\documentclass{aastex631}

\usepackage{savesym}
\usepackage{subfigure,epsfig,url,bm,amsmath,cases,color,soul,longtable,orcidlink}
\savesymbol{tablenum}
\usepackage{siunitx}
\restoresymbol{SIX}{tablenum}

\begin{document}

\title{A Potential Signature of HD 7977's Passage Among Observed Long-Period Comet Orbits}

\author{Nathan A. Kaib\orcidlink{0000-0001-5272-5888}}
\affiliation{Planetary Science Institute, 1700 E. Fort Lowell, Suite 106, Tucson, AZ 85719, USA}
\author{Sean N. Raymond}
\affiliation{Laboratoire d'Astrophysique de Bordeaux, CNRS and Universit\'e de Bordeaux, All\'ee Geoffroy St. Hilaire, 33165 Pessac, France}

\begin{abstract} 

It is generally presumed that the tidal field of the Milky Way's disk is the main perturbation that has driven observed long-period comets (LPCs) from the Oort cloud into the inner solar system. The tide's influence on the Oort cloud should produce a distinct anisotropy in the arguments of perihelion ($\omega$) of dynamically new LPCs with semimajor axes ($a$) over 10$^4$ au. Simulating LPC production dominated by the Galactic tide, we find that observed dynamically new LPCs are more isotropic than expected. Meanwhile, our simulation exhibits much better agreement between simulated and observed ``returning'' LPCs that have made a handful of passages through the inner solar system prior to discovery. The isotropy of new LPCs can be explained if the Oort cloud is much less centrally concentrated than the conventional Oort cloud formation model predicts. However, a second possibility also exists. Additional simulations we perform show that the observed $\omega$ distributions of new and returning LPCs can both be well-replicated if the star HD 7977 passed within $\sim$6000--10000 au of the Sun $\sim$2.5 Myrs ago. In such a scenario, our solar system is still undergoing the latter stages of a comet shower. These simulations imply the modern observed LPC flux is $\sim$twice as high as the longer-term (tide-dominated) rate. This also implies that estimates of the Oort cloud's population should be revised downward by a factor of $\sim$2. Our LPC analysis predicts the upcoming Gaia data release will favor an HD 7977 impact parameter of $\sim$6000--10000 au. 

\end{abstract}


\section{Introduction}

The Oort cloud is the most distant small body reservoir of our solar system, and as the Sun's local galactic environment gravitationally perturbs it, Oort cloud bodies are occasionally driven to low enough perihelia ($q$) that they pass through the inner solar system, where we observe them as long-period comets, or LPCs \citep{oort50}. \citet{oort50} first theorized this cloud's existence through his discovery of the ``Oort spike,'' a relative overabundence of comets with semimajor axes ($a$) beyond $10^4$ au. If such LPCs first entered the inner solar system on smaller semimajor axes before gravitationally scattering off the giant planets, there would be no dynamical reason for the spike to exist. Instead, this spike could be explained if it coincided with the location of the ultimate source region of new LPCs. At such large average distances and weakly bound heliocentric orbits, these bodies are sensitive to external perturbations from the local galactic environment \citep{tre93}. These perturbations drive rapid changes in orbital inclination and eccentricity (and therefore perihelion) allowing them to enter the inner solar system for the first time on weakly bound orbits after spending their prior Gyrs of existence without ever coming close to the inner planets. During their first passage through the inner solar system, random energy kicks from the giant planets are typically strong enough to either eject LPCs forever on hyperbolic orbits or dramatically lower their semimajor axes to more bound orbits ($a<10^4$ au), after which they make subsequent perihelion passages through the inner solar system as ``returning'' LPCs before eventually being ejected or ``fading'' through uncertain physical processes \citep[e.g.,][]{oort51, whip62, weiss79, nes07}. 

When \citet{oort50} first theorized the existence of a distant comet cloud steadily supplying LPCs, he proposed that perturbations from passing field stars are the main mechanism that drives the perihelia of new comets into the inner solar system. However, it was eventually realized that the tidal field of the Milky Way's disk also perturbs the Oort cloud and contributes to LPC production \citep{mormul86, heitre86}. In fact, \citet{heitre86} demonstrated that over long timescales, the Galactic tide's perturbation on the Oort cloud is stronger than the perturbations from typical field star passages. Indeed, the galactic latitudes and longitudes of LPC perihelia hint that the Galactic tide has been the dominant sculptor of the Oort cloud's orbital structure (from an ecliptic disk to a spherical cloud) on timescales comparable to the solar system's age \citep[e.g.][]{hig20, fouch23}. Thus, on average, we should expect the tide to be the main supplier of new LPCs to the inner planetary region ($q<4$ au). 

Under a scenario where the Galactic tide dominates LPC production, LPC orbits should be nearly isotropic but not completely. Whether the tide is driving a comet's perihelion toward or away from the Sun depends on the orientation of the comet's argument of perihelion relative to the Galactic plane ($\omega_G$). If $\sin{2\omega_G}$ is positive, the Galactic tide will drive perihelion Sunward \citep{dqt87}. If this quantity is negative, perihelion will be pulled away from the Sun \citep{fouch06}. (For the remainder of this paper, we use the variable $s_{2\omega}$ to represent $\sin{2\omega_G}$.) In fact, \citet{matliss04} found that the distribution of $s_{2\omega}$ is biased toward positive values for dynamically new LPCs ($a>10^4$ au) as well as for ``young'' returning LPCs with $10^3<a<10^4$ au\footnote{ The term ``dynamically new'' implies that an LPC has never  made a perihelion passage close enough to have been strongly perturbed by the planets prior its discovery, but the magnitude of planetary perturbations is not a binary function that drops to zero beyond some critical perihelion distance \citep{fern81, dybkro15}. We choose the $a>10^4$ au criterion for our dynamically new LPC classification due to its simplicity and prevalent historical usage, but it is an imperfect proxy that does not strictly guarantee an LPC is new to the inner solar system \citep{dybkro25}.} . Members of this latter group of comets are typically LPCs that have passed through the inner solar system one to several times prior to discovery. The planetary energy kicks incurred on these prior perihelion passages are strong enough and numerous enough to remove the LPCs from the Oort spike ($a>10^4$ au), but not numerous enough to drive their semimajor axes to hundreds of au. This limited number of planetary interactions allow this comet group to still retain the Galactic tide's dynamical signature among their arguments of perihelion. In addition, \citet{kaib22} showed that more distant comets with perihelia near or beyond Saturn also have a strong bias toward $s_{2\omega}=1$. 

In spite of these observations, there may still be reason to question the Galactic tide's dominance in the Oort cloud's recent LPC production. Although most stellar perturbations are weaker than the Galactic tide, this is only true on average, and stellar flybys are a random cross-sectional process. Occasionally, a close stellar passage will occur whose Oort cloud perturbations exceed or even dwarf that of the Galactic tide \citep[e.g.,][]{hills81, hut87, dyb02, dyb05, kaib11}. Moreover, the degree of $s_{2\omega}$ anisotropy that the Galactic tide causes among LPC orbits has not been precisely quantified, and it is also not clear if we should expect this anisotropy to disappear completely during brief, stellar-dominated episodes of LPC production. As such, prior detection of $s_{2\omega}$ anisotropy among LPC orbits cannot currently be taken as incontrovertible evidence that the Galactic tide is currently the main supplier of recent LPCs to the inner solar system.

Furthermore, Gaia data have recently revealed that a notably close encounter between our Sun and the Sun-like star HD 7977 may have occurred $\sim$2.47--2.76 Myrs ago \citep{bail22, gaiaDR3, dyb24}. The time elapsed since this encounter is comparable to the typical orbital period of LPCs in the Oort spike, suggesting that its effects on the LPC population could still be felt today. However, the uncertainty on HD 7977's closest approach distance is still quite large. \citet{bail22} found a median minimum approach distance of $\sim$13000 au, but a 5\% probability that the star passed within $\sim$4000 au, and a 5\% probability that it never came closer than $\sim$24000 au. For closest approach distances near the lower uncertainty bound, HD 7977 would have delivered one of the strongest impulses to the Sun in the past Gyr \citep{rick08, kaibquinn09}. In such a scenario, the inner solar system would be subjected to a powerful comet shower, wherein Earth's comet impact probability could temporarily exceed the (typically much larger) asteroid probability \citep{yeocham13}, and the orbital distribution of the Oort cloud would be dramatically reshaped \citep{cao26, dyb24}. Even at the upper end of the approach distances that Gaia data imply, the magnitude of the velocity impulse that HD 7977 delivered to the Sun would only be expected once every 30--40 Myrs in the current solar neighborhood \citep{rick08, kaibquinn09}. However, there may also be an issue of unresolved binarity for HD 7977. The most recent Gaia data release indicates a slightly elevated goodness-of-fit metric (Renormalized Unit Weight Error, or ruwe) for HD 7977 \citep{bail22}. While HD 7977 has no known companion, it may still be unresolved, in which case the minimum encounter distance range derived from Gaia data is unreliable, since it currently fits astrometric data with a single source \citep{bail22, dyb24}. 

Given the possibility of a recent close stellar passage, our work attempts to answer the still-open questions in our understanding of how the orbital distribution of LPCs depends on recent Oort cloud perturbations. The remainder of this paper describes this work. Section \ref{sec:meth} describes our simulations of LPC production, both in the presence and absence of hypothetical HD 7977 passages. Following this, Section \ref{sec:res} first documents the distribution of $s_{2\omega}$ in observed LPCs and then assesses which of our simulations of HD 7977 passages best replicates the observations. Within our favored range of HD 7977 impact parameters from the previous section, Section \ref{sec:dis} discusses how LPC semimajor axes compare with observations, how our assumptions about Oort cloud structure and comet fading influence our results, and how HD 7977 has altered the past and current LPC flux. Finally, Section \ref{sec:con} summarizes our conclusions.

\section{Dynamical Simulation Methods}\label{sec:meth}

To simulate the production of LPCs from the Oort cloud, we use the SCATR N-body code \citep{kaib11b}. This code is largely derived from the SWIFT RMVS3 integration package \citep{levdun94}, but it has the ability to switch from a heliocentric Keplerian drift to a barycentric one when test particles are beyond 300 au from the Sun. In addition, two tiers of integration steps are used: 3600 days in the barycentric frame and 200 days in the heliocentric frame. Given the code's adaptive timestepping and ability to accurately integrate test particles through high eccentricities, it is ideal for modeling the dynamics of LPCs and Oort cloud bodies.

\subsection{TIDE Simulation}

The work presented in this manuscript is based on two sets of simulations. The first one consists of a single simulation we call `TIDE.' In this simulation, the giant planets are initialized on their modern orbits, and 10$^6$ test particles are placed on initial semimajor axes from 3000 to 50000 au with a power-law probability distribution proportional to $a^{-3.35}$. This matches the power-law seen in recent Oort cloud formation models for semimajor axes beyond 20000 au \citep{vok19}. Although \citet{vok19} see a rollover to a less steep ($a^{-2.72}$) power-law interior to 20000 au, we retain the same power-law index for both simplicity and because an early denser galactic environment can result in an Oort cloud more centrally concentrated than that expected from a static galactic environment \citep[e.g.,][]{fern97, bras06, kaib11}. Test particle eccentricities are drawn from a distribution uniform in $e^2$, consistent with isotropic velocities. Although we do not place a formal upper limit on eccentricity, if an orbit's resulting perihelion is below 35 au, the eccentricity is redrawn in order to avoid initially overpopulating the planetary region with comets, since planetary energy kicks remove Oort cloud bodies from these orbits \citep[e.g.,][]{hills81}. In addition, we assign test particle inclinations assuming a distribution uniform in $\cos{i}$. Finally, longitudes of ascending node, arguments of perihelion, and mean anomalies are randomly assigned from uniform distributions between 0 and 360$^{\circ}$. 

With our 10$^6$-particle system initialized, it is integrated under the influence of the gravity of the Sun and giant planets for 2 Gyrs. This integration length is chosen because the modern Oort cloud's bulk structure changes slowly on Gyr-timescales \citep[e.g.][]{dones04, kaib11}, and our initially isotropized cloud is meant to mimic the modern, evolved cloud rather than a cloud in formation \citep[e.g.][]{dqt87, wietre99}.  Our algorithm's adaptive timestepping allows the gravitational effects of the giant planets to be fully resolved when particles are within 300 au of the Sun. Beyond this distance, the planets' perturbations are sampled more coarsely in time, while particles are integrated about the solar system's barycenter. During this integration, the Galactic tide also perturbs our test particles. This tidal force model is dominated by a disk term (but also includes a weaker radial term) that assumes a mean Milky Way local disk density of 0.1 M$_{\odot}$/pc$^3$ \citep{lev01, chak21}. In addition, our simulation is also subjected to hypothetical field star passages. These field stars are drawn from the observed present-day mass function \citep{reid02} and assigned random velocities from distributions whose dispersion depends on spectral category \citep{gar01, rick08}. (See \citet{kaib25} for a detailed description of the stellar encounter algorithm.) Incoming passing stars are started 1 pc from the Sun and integrated until their distance again exceeds 1 pc. Test particles are removed from the simulation if they exceed a heliocentric distance of 1 pc or upon collision with the Sun or planets. To compile LPC orbital statistics, when particles make perihelion passages inside 20 au, their orbital elements are recorded at a heliocentric distance of 50 au on their inbound leg. In order to avoid prohibitive file sizes, particle orbits are only recorded for semimajor axes over 100 au. 

The main purpose of our TIDE simulation is to model the LPC orbital distribution expected from an epoch of the solar system when the Galactic tide dominates Oort cloud perturbations. Furthermore, we wish to compare this distribution with other distributions that result when the solar system has recently been subjected to HD 7977 passages of varying strength. Thus, for the last Gyr of our TIDE simulation, the only perturbations we include on the Oort cloud are our tidal perturbation model and ``weak'' stellar passages. To still include a weak stellar passage component, we manually prevent any stellar passages from occurring in the simulation's second Gyr that would yield greater impulses or impulse gradients \citep{kaibray24} to the Sun than would be generated if HD 7977 passed within 30000 au. This process still allows stellar passages to influence the Oort cloud, but the actual passage events will be minor compared to the hypothetical HD 7977 passages we will eventually consider, and it also generates a tide-dominated set of LPC orbits during this simulation's final Gyr. The reason we take care to still include a weak stellar passage perturbation component is that it has a continual isotropizing influence on the overall Oort cloud orbital distribution, and it prevents the Galactic tide from exhausting its LPC-producing regions of phase space in the Oort cloud \citep{rick08}. 

\subsection{Passage Simulations}

To complement our TIDE simulations, we also run a set of 13 shorter simulations designed to model the LPC population seen 2--3 Myrs after hypothetical HD 7977 passages. Because of the short time-of-interest of these simulations, they have a much shorter integration time of just 3.5 Myrs. However, unlike the TIDE simulation, they must resolve changes to the LPC orbital distribution as they occur on $\sim$Myr timescales. During tide-dominated epochs, the rate that the Oort cloud generates new LPCs in the inner solar system ($q<4$ au) is roughly 1 LPC per 10$^5$ Oort cloud bodies per Myr \citep{kaibquinn09}. Thus, a $\sim$10$^6$-particle simulation like TIDE will only generate $\sim$10 LPCs during a 1-Myr interval. To increase the number of particles in these new simulations to $\sim$10$^7$ (which would yield $\sim$100 new LPCs per Myr), we co-add ten different time snapshots of the particles in our TIDE simulation. These snapshots are acquired every 100 Myrs between $t=1.1$ Gyrs through $t=2$ Gyrs in our TIDE simulation. Since the second Gyr of the TIDE simulation does not contain unusually powerful stellar passages, the dynamical state of the Oort cloud and LPC distributions in our snapshots should all be roughly statistically similar. Thus, our second set of simulations all begin with the same $\sim$10$^7$-particle Oort cloud whose only recent external perturbations have been the Galactic tide and weak stellar passages. Just like the TIDE simulation, these simulations also include the Sun and the four giant planets on their modern orbits. 

Each of our passage simulations includes a single stellar encounter representing HD 7977. Gaia Data Release 3 (DR3) lists this star's radial velocity as 26.8 km/s \citep{gaiaDR3}. The current Right Ascension and Declination of HD 7977 are listed as \SI{1}{\hour} \SI{20}{\minute} \SI{31.596}{\second} and +61.8825$^{\circ}$, respectively. In addition, Gaia DR3 lists a nominal R.A. proper motion of 0.144 mas/yr and a nominal Declination proper motion of 0.01 mas/yr. This current position and velocity data constitute an ``exit point'' for each passage simulation's stellar passage. To get an entry point and entry velocity vector for the beginning of each of our stellar passages, HD 7977's currently observed motion is run backwards in time past its minimum distance from the Sun until it again reaches a distance of 1 pc with an incoming velocity vector (when run forward in time). Transforming the entry position vector to a galactic frame of reference using Astropy, we attain an initial starting point for the stellar passage in each of our passage simulations that is 1 pc from the Sun. The initial radial velocity of each simulation's stellar passage is set to -26.8 km/s, and the initial tangential velocity direction is derived from the Gaia DR3 proper motions (transformed to the galactic frame), but the magnitude of the tangential velocity is scaled differently in each of our simulations to achieve different impact parameters for HD 7977. (We later became aware of a faster radial velocity measurement for HD 7977 of 29.9 km/s \citep{err20}, but given the modest difference in approach speed and resulting impulses, we did not rerun all of our simulations.)

Each passage simulation's stellar encounter is initiated at $t=0.1$ Myr, and simulations are integrated for a total of 3.5 Myrs. Our 13 different simulations explore 13 different impact parameters for HD 7977. They are spaced logarithmically from 3162 au (10$^{3.5}$ au) to 50119 au (10$^{4.7}$ au) in increments of 0.1 dex. The naming scheme of these passage simulations is `Flyby\_$<$X$>$,' where $<$X$>$ is the HD 7977 impact parameter measured in au. Thus, the simulation with the most powerful HD 7977 passage is named Flyby\_3162 and the one with the weakest passage is Flyby\_50119. As in the TIDE simulation, when particles make perihelion passages inside 20 au, their orbital elements are recorded at a heliocentric distance of 50 au on their inbound leg. 

\section{Results}\label{sec:res}

\subsection{Observed Long-Period Comets}

To evaluate our simulations' results, we compare the orbits of simulated LPCs with those of observed LPCs. There now exist hundreds of known LPCs, and they have primarily been discovered over the last 150 years. Incredibly, a sample of just 19 observed LPCs proved sufficient for \citet{oort50} to infer the existence of the Oort cloud. However, we are interested in potentially subtle variations in the distribution of $s_{2\omega}$ for LPCs, which are linked to the spatial orientation of LPC orbits. To compare these variations in our simulations with observations, we require large numbers of observed LPCs whose orbits are known to be accurate, and we also require them from eras of observation in which searches for new comets extended across the entire celestial sphere. It is likely that comet discoveries in the first half of the 1900s and earlier do not meet this requirement. \citet{fran05} notes that prior to 1990, annual LPC discoveries are dominated by amateur astronomers, but after this point, systematic sky surveys discover a larger and larger fraction of LPCs. Based on this, we require our observed LPCs to be discovered after 1989 when comparing them against simulations. 

\begin{figure}
\centering
\includegraphics[scale=0.55]{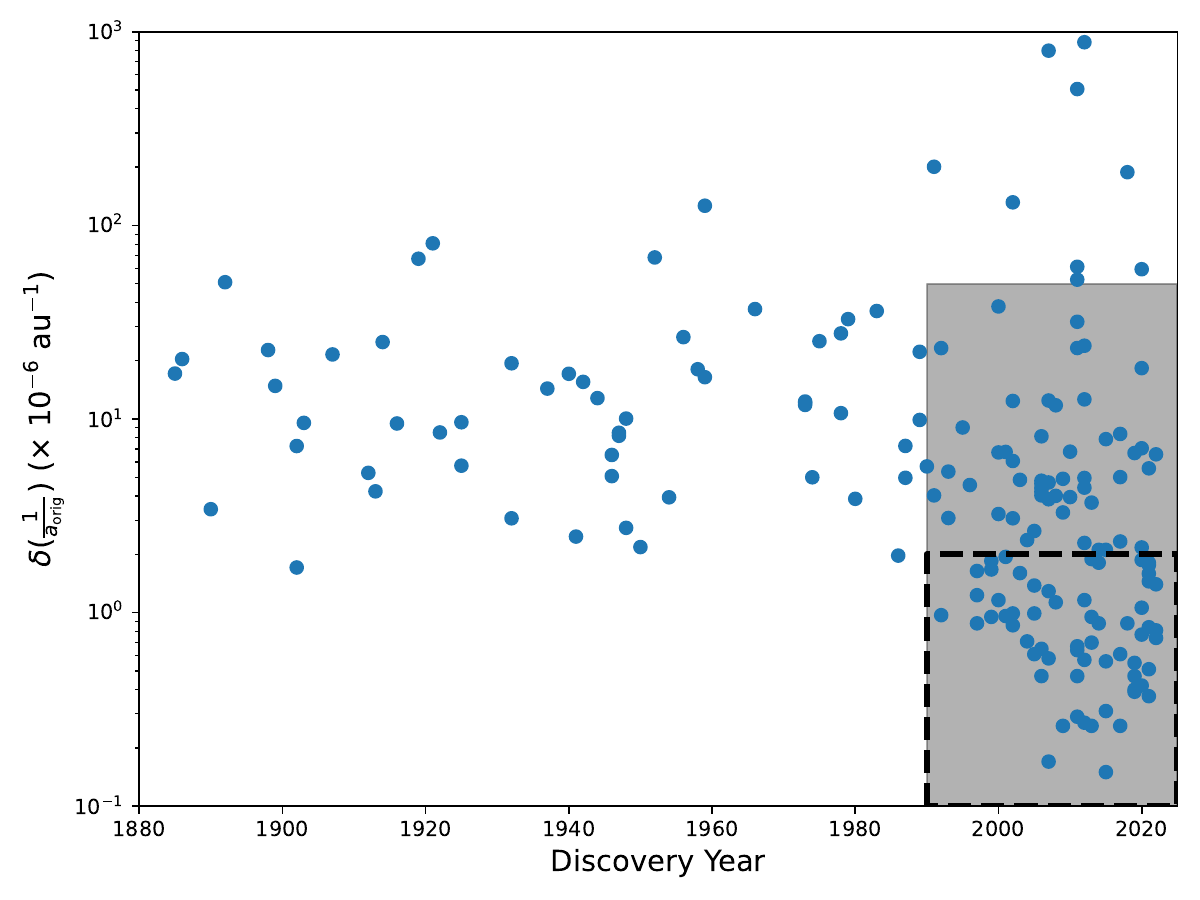}
\caption{Estimated uncertainty of LPC original semimajor axis vs LPC discovery year for dynamically new ($\frac{1}{a_{orig}} < 10^{-4}$ au$^{-1}$) LPCs listed in the CODE catalogue \citep{krodyb20}. The shaded rectangle marks the time range and uncertainty range containing the observed dynamically new LPCs that we consider in most of our paper's analyses. The dotted line marks all dynamically new LPCs discovered since 1990 with original semimajor axis uncertainties below $2\times10^{-6}$ au$^{-1}$.}
\label{fig:obstrends}
\end{figure}

It is also important to consider the uncertainty in an observed LPC's original orbit (the orbit it possesses on its inbound trajectory before it receives significant planetary energy kicks near perihelion), as this is the basis of whether it is classified as dynamically new ($a>10^4$ au) or returning ($a<10^4$ au) \citep[e.g.][]{fern81}. To evaluate these uncertainties for individual comets, \citet{krodyb20} backintegrate swarms of particles distributed on orbits consistent with that particular LPC's observations and then measure the swarm's distribution of barycentric orbital elements at large heliocentric distance \citep{kro01, sit98}. These uncertainties have been calculated for virtually every dynamically new LPC ($a>10^4$ au) discovered before 2023 and are listed in the CODE catalog \citep{krodyb20, krodyb23, dybkro25}. In Figure \ref{fig:obstrends}, we plot the uncertainty in the inverse of the original semimajor axis for every dynamically new LPC in the CODE catalog against LPC discovery year. (Figure \ref{fig:obstrends} uses the CODE catalog's ``preferred orbits,'' which may correct a comet's inferred original orbit for non-gravitational accelerations when these forces are well-characterized for that particular comet. Later, in Figure \ref{fig:smacompare}A, we also consider the original orbits of CODE LPCs assuming pure gravitational fitting.) Here we see that there is a marked improvement in the number of precisely known original LPC orbits beginning around 1990. This coincides with the increasing contribution of sky surveys toward LPC discoveries \citep{fran05}. In the remaining sections of our paper, we often split our simulated LPCs into two groups: dynamically new comets ($a>10^4$ au) and young returning comets ($10^3<a<10^4$ au). To avoid comparing our simulations against too many observed LPCs that are misclassified among those two groups, we therefore require our observed dynamically new LPCs to have original semimajor axis uncertainties below $50\times10^{-6}$ au$^{-1}$, or half the orbital energy of a cometary semimajor axis of 10$^4$ au. The dynamically new LPCs that we compare our simulations against are shown in the shaded region of Figure \ref{fig:obstrends} and consist of 107 observed comets. 

As mentioned above, we also wish to compare our simulation's production of young returning LPCs against those that are observed. Unfortunately, the CODE catalog's orbital analysis does not extend to most known returning LPCs. To build a sample of observed young returning LPCs, we therefore utilize the comets listed in the JPL Horizons system. This catalog lists nearly 700 returning LPCs. Based on Figure \ref{fig:obstrends}, we assume that the observational trends in the discoveries of returning LPCs mirror those seen in new LPCs, and we consequently only consider returning LPCs discovered after 1989 and prior to 2023 (the year before which the CODE catalog is nearly complete among Oort spike comets). To compute these comets' original semimajor axes, we calculate their barycentric orbital elements on January 01 1950 using the Horizons ephemerides generating system. The observed LPCs with original semimajor axes between $10^{3-4}$ au comprise our observed sample of young returning LPCs against which we compare simulation results. In total, these criteria yield 112 observed young returning LPCs. 

\subsection{Overly Isotropic New Long-Period Comets}

We begin our comparison between simulated LPCs and observed ones using our TIDE simulation. In this simulation, we consider LPCs that make perihelion passages inside $q<4$ au during the last 900 Myrs, which are temporally well-removed from the simulation's first half of evolution that included powerful stellar passages. During this timespan, we record 5677 instances of 5016 unique simulated bodies making perihelion passages on orbits with $a>10^4$ au, and these represent our sample of simulated dynamically new LPCs. In addition, there are 13130 perihelion passages of 2382 unique bodies with semimajor axes between $10^3$ and $10^4$ au, which comprise our simulated sample of young returning LPCs. It is obvious from the above numbers that particles sometimes make multiple perihelion passages through the inner solar system to contribute to our orbital samples multiple times. To account for the fact that LPCs seem to rapidly fade (and have lower discovery probability) through still-debated physical processes during sequential perihelion passages, we weight a particle's contribution according to the number of times it has passed through the inner solar system, which we define as $q<4$ au \citep{oort51}. We do this using Whipple's fading scheme \citep{whip62}, which assumes the probability that a comet will survive at least $n$ perihelion passages is given by 
\begin{equation}
\Phi = n^{-\kappa}
\label{eq:whip}
\end{equation}
where $\kappa$ is simply a fitted power-law index. We do not perform a thorough evaluation of different $\kappa$ values but instead rely on prior work, which has found that $\kappa\simeq0.6$ provides reasonable simulated matches to the observed ratio of new LPCs to returning LPCs \citep{wietre99, vok19}. 

\begin{figure}
\centering
\includegraphics[scale=0.55]{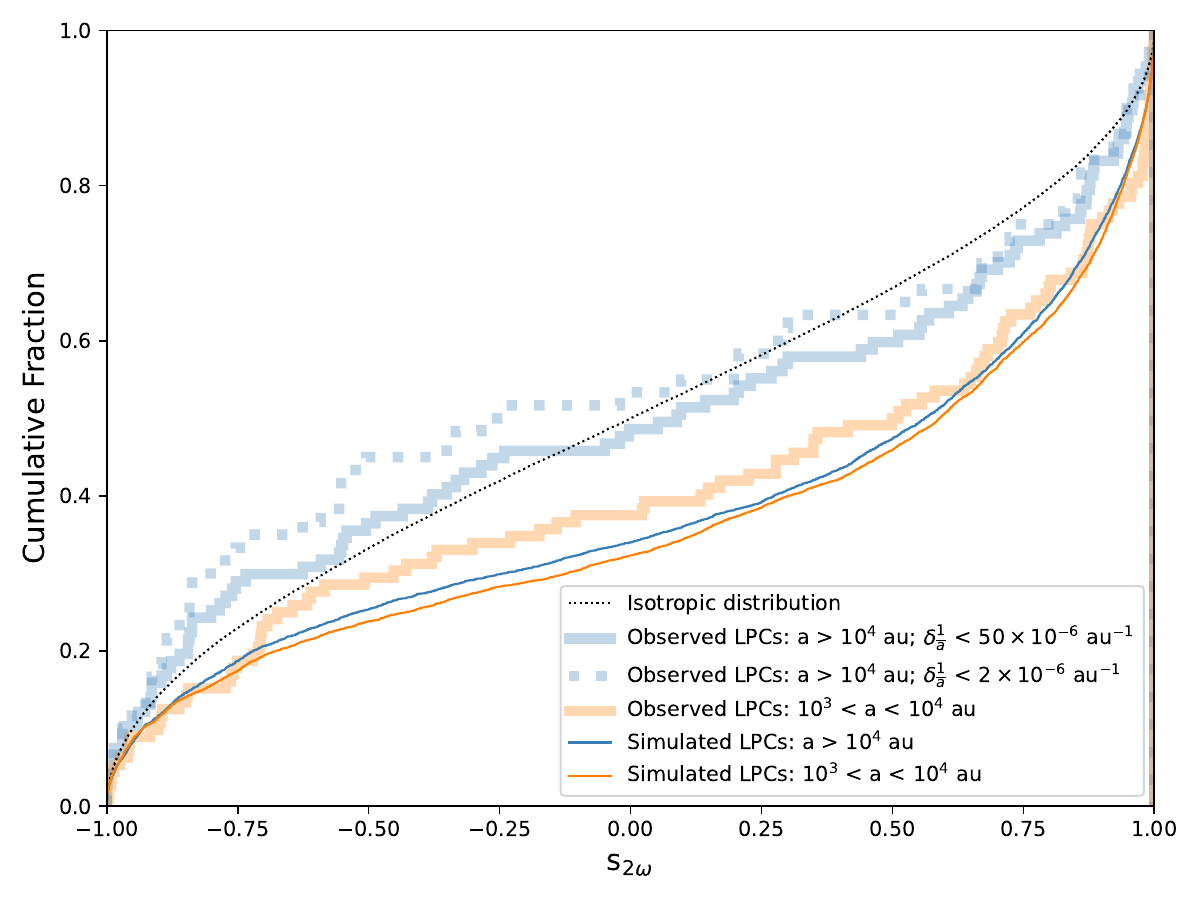}
\caption{The cumulative distribution of $s_{2\omega}$ is shown for observed LPCs (thick lines) and simulated LPCs from the TIDE simulation (thin lines). LPCs are split into dynamically new comets ($a>10^4$ au; {\it blue}) and returning comets with $10^3<a<10^4$ au ({\it orange}). Among observed LPCs, we only consider comets discovered since 1990, and we also require uncertainties below $50\times10^{-6}$ au$^{-1}$ in the original semimajor axes of our dynamically new LPCs \citep{krodyb19}. The dashed blue line is for observed dynamically new LPCs with original semimajor axis uncertainties below $2\times10^{-6}$ au$^{-1}$. The dotted line marks a completely isotropic distribution of orbit orientations.}
\label{fig:obsdiscrep}
\end{figure}

The distribution of $s_{2\omega}$ for each of our two groups of simulated LPCs is shown in Figure \ref{fig:obsdiscrep}. One can see that neither group is isotropically distributed. In both groups, only $\sim$1/3 of orbits are found with $s_{2\omega}<0$. Moreover, the simulated distributions are very similar to one another, which is unsurprising, since returning LPCs are derived from new LPCs. The simulated young returning LPCs are slightly less isotropic than the dynamically new LPCs. This is likely because a small number of LPCs with $s_{2\omega}<0$ will have their perihelia pulled out of the $q<4$ au region by the Galactic tide if their semimajor axes are just under $10^4$ au and their prior perihelia are slightly under 4 au. 

The $s_{2\omega}$ distributions for observed LPCs are also shown in Figure \ref{fig:obsdiscrep}. We see that the observed population of young returning LPCs resembles the distribution seen in our TIDE simulation. When we use a Kolmogorov-Smirnov (K-S) test to compare the $s_{2\omega}$ distribution of our simulated young returning LPCs with that of the observed ones, we get a $p$-value of 0.57, indicating that we cannot reject the null hypothesis that the two samples have the same underlying distribution\footnote{Elements of our sample of simulated LPC orbits have different weights due to the Whipple fading law we implement. To estimate $p$-values of statistical comparison tests, we resample the simulated orbital list in a bootstrap manner where each element's resampling probability is proportional to its Whipple survival probability. Then the statistical test is performed on this bootstrapped sample instead of the original. This process is performed 100 times and the median returned $p$-value is selected as the $p$-value we quote in the text.}. However, the situation is very different when we look at the observed $s_{2\omega}$ distribution of dynamically new LPCs. Nearly half (48.6\%) of observed dynamically new LPCs have $s_{2\omega}<0$. When we compare this sample against the simulated sample, a K-S test returns a $p$-value of 0.0039. (Similarly, an Anderson-Darling test returns a $p$-value of 0.0035.) This suggests we can reject the null hypothesis that the two distributions are drawn from the same parent distribution with nearly 3$\sigma$ confidence. In fact, the observed $s_{2\omega}$ distribution for dynamically new LPCs is much more consistent with an isotropic distribution, and a K-S test returns a $p$-value of 0.47 in that instance. This discrepancy between the TIDE simulation and observed dynamically new LPCs is not resolved if we only consider LPCs with more precisely known orbits. If we decrease our tolerance in original semimajor axis uncertainty from $<50\times10^{-6}$ au$^{-1}$ down to $<2\times10^{-6}$ au$^{-1}$, we see in Figure \ref{fig:obsdiscrep} that there is no improvement to the match to the TIDE simulations. This smaller subset of orbits represent the highest quality tier of observed LPC orbits \citep{krodyb19}. Comparing this more precise set of observed dynamically new LPCs against the TIDE simulation, we find that the $p$-values are essentially unchanged, as the smaller observed sample size is offset by a still larger apparent disagreement with the simulation.

The significant bias toward positive $s_{2\omega}$ in our TIDE simulation's LPCs in Figure \ref{fig:obsdiscrep} occurs due to timescale disparities. Since $s_{2\omega}$ determines the direction of LPC perihelion change under the Galactic tide, a comet must always have $s_{2\omega}>0$ when its perihelion first crosses from $q>4$ au to $q<4$ au. To be observed as an LPC with $s_{2\omega}<0$, after evolving across the $q=4$ au boundary, the comet's perihelion must reach its minimum value and then begin evolving outward from the Sun again before making an actual perihelion passage (when it would be observed as an LPC). Alternatively, if it were to be observed as an LPC with $s_{2\omega}>0$, the comet would simply make a perihelion passage before its perihelion reached the minimum value during its evolution under the Galactic tide. For a given comet/orbit, the former perihelion evolution obviously requires more time to complete than the latter. Meanwhile, the time between actual perihelion passages is simply dependent on a comet's orbital period. Thus, more dynamically new LPCs are observed with $s_{2\omega}>0$ than with $s_{2\omega}<0$. Since the planetary energy kick incurred during a dynamically new LPC's perihelion passage is typically greater than its orbital energy \citep{fern81}, there is no further evolution of the $s_{2\omega}$ distribution for the group of LPCs with $a>10^4$ au (whose observed orbits are classified as dynamically new). For those dynamically new LPCs that incur a negative planetary energy kick during perihelion passage, most will make their next perihelion passage as a ``young'' returning LPC with $10^3<a<10^4$ au. With smaller semimajor axes, these young returners evolve much more slowly under the Galactic tide compared to dynamically new LPCs \citep{heitre86}. Consequently, the values of perihelion and $s_{2\omega}$ will change very little for young returners compared to their original values they possessed on their first perihelion passages as dynamically new LPCs. This is why the TIDE simulation's $s_{2\omega}$ distribution for young returning LPCs looks so similar to the simulation's dynamically new distribution in Figure \ref{fig:obsdiscrep}, and the observed differences in the $s_{2\omega}$ distributions for young returning and dynamically new LPCs cannot be replicated with the TIDE simulation's dynamics.

\begin{figure}
\centering
\includegraphics[scale=0.55]{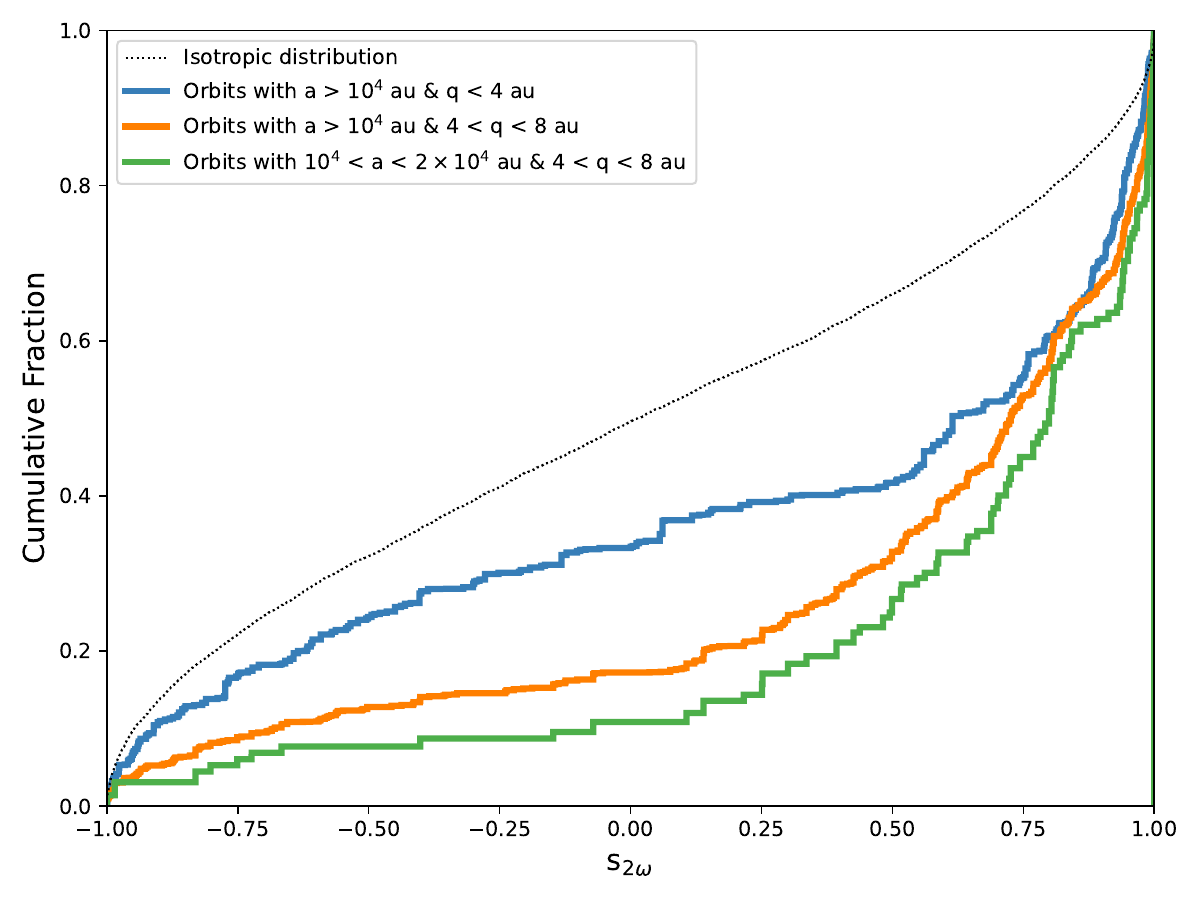}
\caption{The distribution of $s_{2\omega}$ for Oort cloud orbits in ten co-added snapshots of our TIDE simulation ($t=1.1$ to $t=2.0$ Gyrs in 100-Myr increments). Distributions are shown for orbits with $q<4$ au and $a>10^4$ au ({\it blue}), orbits with $4<q<8$ au and $a>10^4$ au ({\it orange}), and orbits with $4<q<8$ au and $10^4<a<2\times10^4$ au ({\it green}). Each orbit is weighted by its orbital frequency ($a^{-1.5}$). The dotted line marks an isotropic distribution.}
\label{fig:s2omegaasym}
\end{figure}

In the absence of planetary energy kicks, we would expect the $s_{2\omega}$ distribution to be isotropic for LPCs with $a>10^4$ au and $q<4$ au, since their $s_{2\omega}$ values can continue to evolve after they make their first perihelion passage through the inner solar system. One therefore might naively expect the $s_{2\omega}$ anisotropy in our TIDE simulation to decrease as we move away from the Sun and closer to the edge of the LPC loss cone \citep{hills81, heitre86}. Instead, Figure \ref{fig:s2omegaasym} shows that the opposite occurs. In this figure, we sample the orbital distribution of the TIDE simulation's Oort cloud 10 times: once every 100 Myrs between $t=1.1$ Gyrs and $t=2$ Gyrs. Then we co-add the snapshots to build a large set of low-$q$ orbits in a scenario where the Galactic tide dominates Oort cloud dynamics. In Figure \ref{fig:s2omegaasym}, we plot the $s_{2\omega}$ distribution for orbits with $a>10^4$ au and $q<4$ au (weighted by inverse orbital period to account for relative contribution to LPC flux). In this case, we see almost exact agreement with the $s_{2\omega}$ distribution gathered from the perihelion passages of dynamically new LPCs in the TIDE simulation; 1/3 of orbits have $s_{2\omega}<0$. We then look at the distribution of $s_{2\omega}$ for TIDE simulation orbits with $a>10^4$ au and $4<q<8$ au. Here we see that for these LPCs with perihelia just outside the inner solar system, the bias toward $s_{2\omega}=1$ is even stronger than in the inner solar system. The reason is that LPCs can only enter the $4<q<8$ au zone from the $q>8$ au region, since LPCs making perihelion passages inside 4 au will be ejected or placed onto more tightly bound orbits after their first low-$q$ passage. This requires LPCs to have $s_{2\omega}>0$ when their perihelion first enters the 4--8 au region, since they can only enter from higher values of $q$. To reverse the sign of their $s_{2\omega}$ value (in the tide-dominated dynamical regime), these LPCs' perihelia must evolve to their minimum perihelion and then begin moving outward from the Sun again prior to perihelion passage in the $4<q<8$ au region. For many of these comets, this can require even longer timescales than for LPCs in the $q<4$ au region, since the minimum perihelion value may be below 4 au. Thus, some LPCs' perihelia must evolve from beyond 8 au to interior to 4 au and then back outward to 4--8 au in less than a single revolution around the Sun. Consequently, there is even stronger anisotropy for dynamically new LPCs in the $4<q<8$ au region even though they are not as deep in the Oort cloud's loss cone. 

In Figure \ref{fig:s2omegaasym}, we also plot the $s_{2\omega}$ distribution for orbits with $4<q<8$ au and $10^4 < a < 2\times10^4$ au. Orbits with this range of semimajor axes do not typically evolve quickly enough under the Galactic tide to penetrate the inner solar system, but they can reach the Jupiter-Saturn region of the loss cone. Here we see that their $s_{2\omega}$ distribution is even more extreme than the $a>10^4$ au sample of orbits. Restricting the maximum semimajor axis to $2\times10^4$ au means that the typical orbital period is shorter, and therefore less time is allotted for these orbits to reverse their $s_{2\omega}$ sign in a single revolution. In addition, as semimajor axis decreases, orbital evolution under the Galactic tide becomes slower, and the typical time required for an orbit to reverse $s_{2\omega}$ sign increases. Both of these effects are what drives this $s_{2\omega}$ distribution still closer to 1. 

\subsection{The Effects of HD 7977 Passages}

The weakness of anisotropy among observed new LPCs suggests that the Galactic tide may not have been solely responsible for the recent production of LPCs. Perhaps the most plausible source of additional recent significant perturbation to the Oort cloud is HD 7977, which could have passed as close as 4000 au to the Sun $\sim$2.47--2.76 Myrs ago \citep{bail22, dyb24}. For this reason, we now explore the LPC orbits generated in our second set of simulations that modeled different hypothetical HD 7977 passages. 

\begin{figure}
\centering
\includegraphics[scale=0.32]{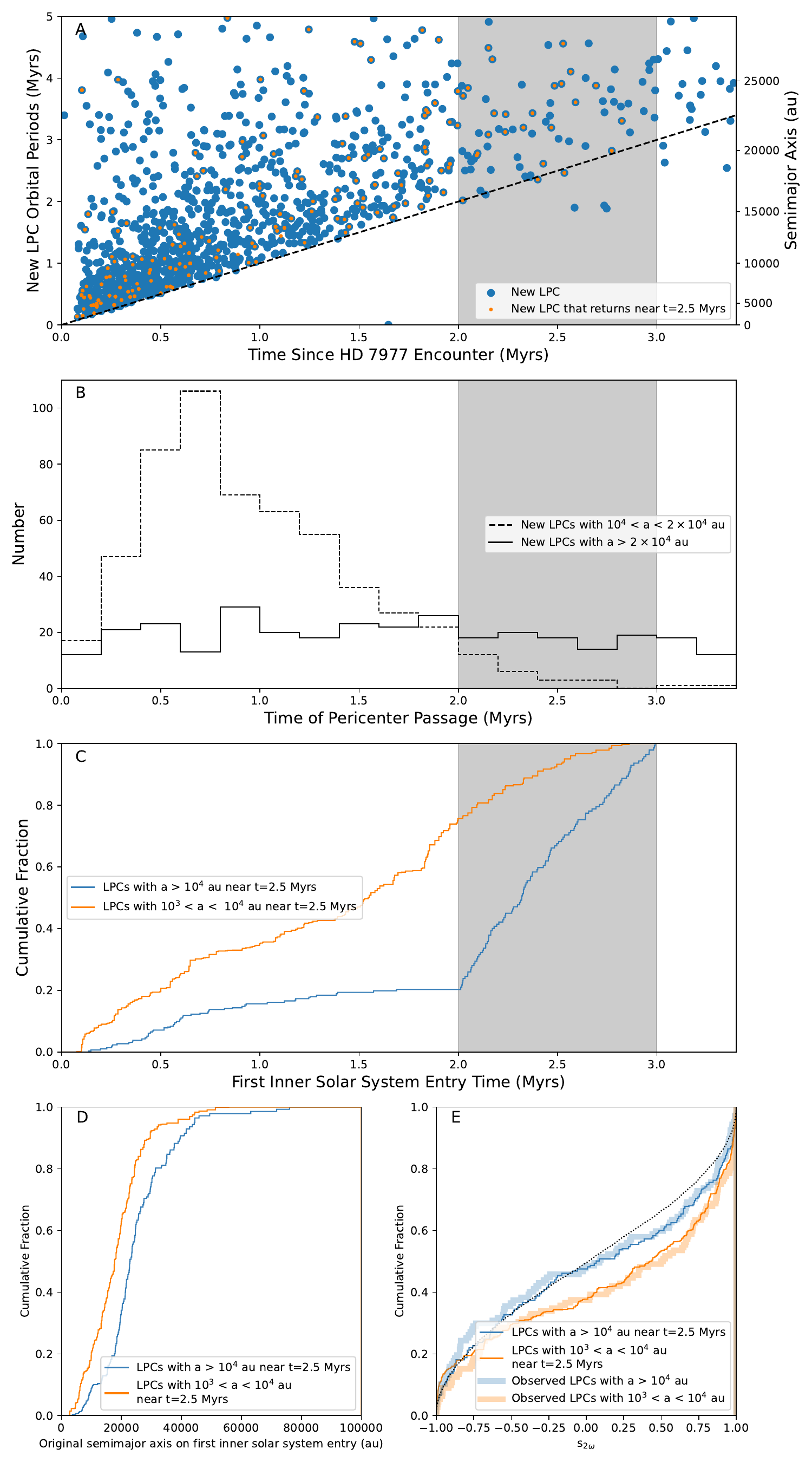}
\caption{{\bf A:} Plot of LPC orbital period vs perihelion passage time for LPCs making their first perihelion passage inside 4 au in the Flyby\_7943 simulation. LPCs overplotted in orange go on to make returning passages inside 4 au with $10^3<a<10^4$ au at 2--3 Myrs after HD 7977's passage (time range shaded in gray). {\bf B:} Distribution of first perihelion passage times inside 4 au. Time distributions are shown for LPCs with $a>2\times10^4$ au ({\it solid}) and with $10^4<a<2\times10^4$ au ({\it dashed}). {\bf C:} For LPCs making perihelion passages 2--3 Myrs after HD 7977's passage, the distribution of times at which they make their first inner solar system perihelion passage is shown. Distributions are shown for LPCs with $a>10^4$ au ({\it blue}) and with $10^3<a<10^4$ au ({\it orange}). LPC contributions to the distributions are weighted with Equation \ref{eq:whip}. {\bf D:} For LPCs making perihelion passages 2--3 Myrs after HD 7977's passage, the distribution of the semimajor axes they possess on their first perihelion passage through the inner solar system is shown. Distributions are shown for LPCs with $a>10^4$ au ({\it blue}) and with $10^3<a<10^4$ au ({\it orange}). LPC contributions to the distributions are weighted with Equation \ref{eq:whip}. {\bf E:} For LPCs making perihelion passages 2--3 Myrs after HD 7977's passage, their $s_{2\omega}$ distributions they have on these perihelion passages is shown. Distributions are shown for LPCs with $a>10^4$ au ({\it blue}) and with $10^3<a<10^4$ au ({\it orange}). LPC contributions to the distributions are weighted with Equation \ref{eq:whip}. }
\label{fig:orbperiod}
\end{figure}

In Figure \ref{fig:orbperiod}, we show the results of our Flyby\_7943 simulation, in which HD 7977 passes the Sun with an impact parameter of 7943 au. In Panel A, we plot the orbital periods (and semimajor axes) of LPCs making their first passage through the inner 4 au of the solar system against the time of their perihelion passage. (Unlike for observed LPCs, we do not rely on the $a>10^4$ au criterion to classify a simulated LPC as new in this instance. In simulations, we can know particles' exact orbital histories.) The effects of the HD 7977 passage are obvious and dramatic. Beginning about 10$^5$ years after HD 7977's passage, the number of LPCs increases markedly. There is also a direct relationship between the minimum semimajor axis an LPC can have on its first perihelion passage and the time elapsed since the HD 7977 encounter. The orbital period of a new, HD 7977-triggered LPC is almost never shorter than the time elapsed since the HD 7977 flyby. Since our Oort cloud is centrally concentrated and relative LPC flux also scales with orbital frequency ($a^{-1.5}$), the new LPC semimajor axis distribution at any given time after the HD 7977 encounter is weighted toward the minimum value allowed. In this simulation, many LPCs make their first passage through the inner solar system on semimajor axes smaller than any seen in our TIDE simulation. Immediately after the stellar passage, most new LPCs have $a\sim5000$ au. However, by 1 Myr after the passage, the typical new LPC semimajor axis increases to $\sim$10$^4$ au. Finally, 2--3 Myrs after the HD 7977 encounter new LPC semimajor axes are near $\sim$$2\times10^4$ au. 

In Figure \ref{fig:orbperiod}A, we also note new LPCs in our Flyby\_7943 simulation that go on to make perihelion passages on orbits that would be classified as young returning LPCs during the time window 2--3 Myrs after the HD 7977 passage. We take this time window as being analogous to the modern epoch of the solar system if it was strongly perturbed by HD 7977 $\sim$2.5 Myrs in the past. Here we can see that many of the new LPCs that go on to become returning LPCs first pass through the inner solar system well before the start of the 2--3 Myr time window. Because the typical semimajor axis of a new LPC is growing with time, this means that many young returning LPCs passing through the inner solar system in the 2--3 Myr period are derived from an orbital distribution of new LPCs that is different from the new LPCs seen in the 2--3 Myr window.

Figure \ref{fig:orbperiod}B confirms this. Here we plot the distribution of pericenter passage times for simulated new LPCs with semimajor axes between 1--2 $\times10^4$ au as well as for new LPCs with semimajor axes over $2\times10^4$ au. We find that until 2 Myrs after the HD 7977 passage, there are more new LPCs with semimajor axes under $2\times10^4$ au than those over this value. In fact, the group of LPCs on smaller semimajor axes is 2--3 times more common than LPCs with $a>2\times10^4$ au for much of the first 1.5 Myrs after HD 7977's passage. After 2 Myrs, the situation reverses, and LPCs with $a>2\times10^4$ au become more common, and nearly no new LPCs have semimajor axes under $2\times10^4$ au after $t=2.5$ Myrs. (In the TIDE simulation, LPCs with $10^4 < a < 2\times10^4$ au only rarely pass through the inner solar system.)  

For the simulated LPCs making pericenter passages between 2--3 Myrs after HD 7977, we next examine the times these bodies first make a passage through the inner solar system ($q<4$ au) in Figure \ref{fig:orbperiod}C. We split these LPCs into two groups: 1) LPCs that would be observationally classified as new ($a>10^4$ au) and 2) LPCs that would be considered young returners ($10^3<a<10^4$ au). For the first group, we indeed find that most of them (80\%) first pass through the solar system between 2 and 3 Myrs after the HD 7977 passage, justifying the $a>10^4$ au criterion as a reasonably accurate criterion for dynamically new LPCs. Surprisingly, however, there is also a minor contribution from LPCs entering the inner solar system before the 2--3 Myr window, and they are primarily from the first Myr of the simulation. (We weight simulated LPCs' contributions to distributions with their Whipple survival probability given in Equation 1, where $\kappa=0.6$.) These are mostly LPCs that are originally injected into the inner solar system with semimajor axes between $\sim$5000 and $\sim$10$^4$ au. These LPCs survive their first passage through the inner solar system, receive positive planetary energy kicks, and later return as LPCs with $a>10^4$ au during the 2--3 Myr time window. Nevertheless, the typical LPC with $a>10^4$ au during our time window of interest is on its first passage through the inner solar system, and the median entry time into the inner solar system for these LPCs is $\sim$2.3 Myrs after the HD 7977 passage.

Figure \ref{fig:orbperiod}C shows a quite different situation for LPCs with $10^3<a<10^4$ au between 2--3 Myrs after the HD 7977 passage. Roughly 75\% of these LPCs first enter the solar system before 2 Myrs, and the median time of entry is $\sim$1.5 Myrs after the HD 7977 passage, which is $\sim$0.8 Myrs earlier than our $a>10^4$ au LPCs. As Figure \ref{fig:orbperiod}A shows, the LPC population changes dramatically over this $\sim$Myr timescale. Figure \ref{fig:orbperiod}D confirms that for the 2--3 Myr window, LPCs classified as young returners and LPCs classified as dynamically new are sampling different regions of the Oort cloud. On their first inbound leg into the inner solar system, the original semimajor axis of the young returners has a median of 17700 au. This is over 5000 au smaller than the median original semimajor axis of the dynamically new LPCs, which is $\sim$22900 au. Moreover, nearly 40\% of young returners originate from the inner 15000 au of the Oort cloud, whereas this region only supplies $\sim$13\% of the dynamically new LPCs. 

As seen in Figure \ref{fig:s2omegaasym}, the anisotropy of the $s_{2\omega}$ distribution depends sensitively on semimajor axis for bodies with perihelia just outside the inner solar system ($4<q<8$ au). In the event of a strong perturbation to the perihelia of Oort cloud orbits during HD 7977's passage, these are the bodies most likely to be newly injected into the inner solar system. A stellar perturbation powerful enough to perturb these bodies' perihelia sunward by $\sim$4 au into the inner solar system may not necessarily be powerful enough to completely isotropize their $s_{2\omega}$ distribution. As a result, in the aftermath of an HD 7977 passage, young returning LPCs may exhibit less $s_{2\omega}$ isotropy than dynamically new LPCs on account of the returners being derived from more interior Oort cloud orbits than dynamically new LPCs. Figure \ref{fig:orbperiod}E indicates that this is exactly what is seen in the Flyby\_7943 simulation. For young returning LPCs in the 2--3 Myrs after HD 7977, 38\% have negative values of $s_{2\omega}$, but for dynamically new LPCs, this fraction increases to 48\%. In addition, we see that both simulated groups of LPCs have $s_{2\omega}$ distributions that closely resemble their observed counterparts. 

\begin{figure}
\centering
\includegraphics[scale=0.4]{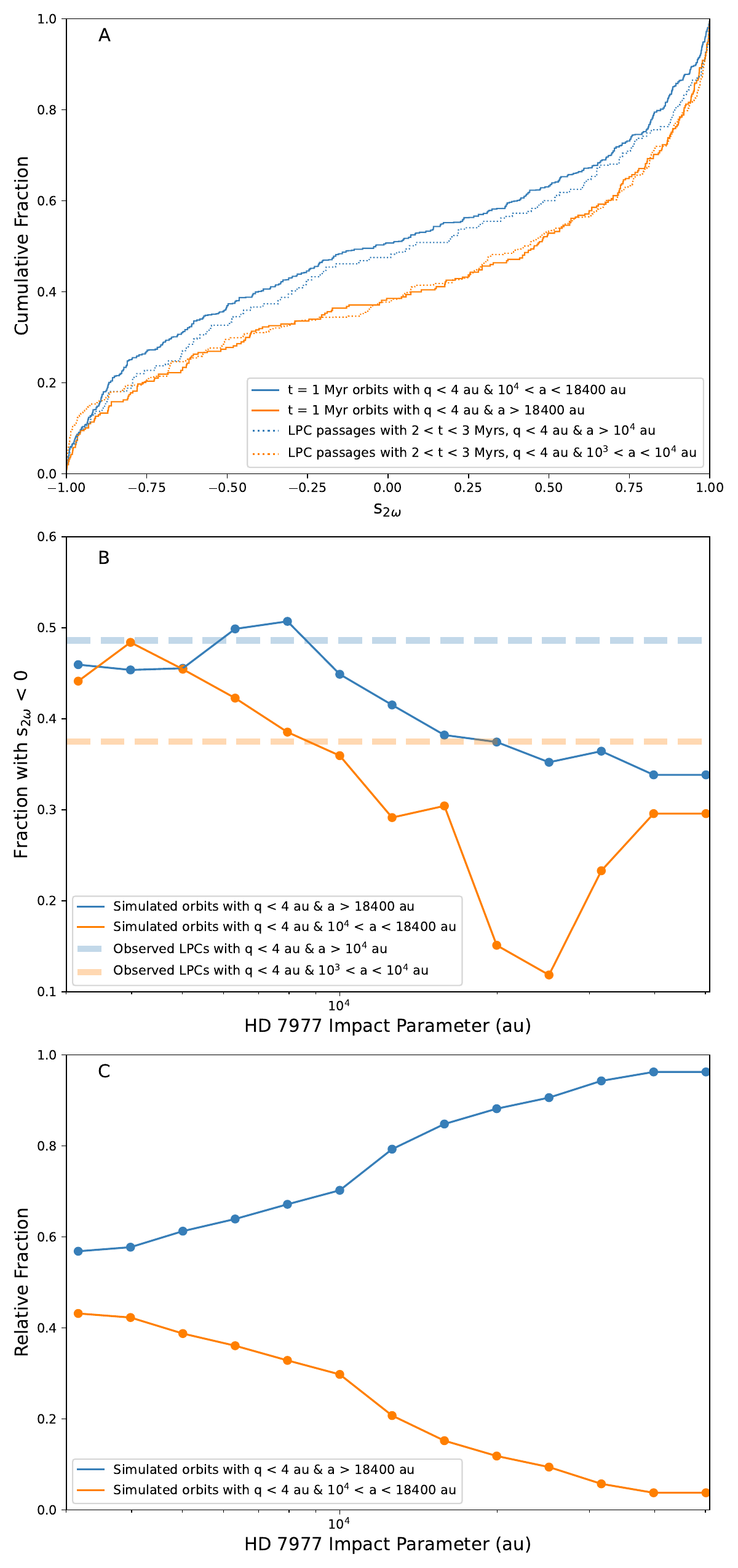}
\caption{{\bf A:} For the Flyby\_7943 simulation, the distribution of $s_{2\omega}$ at $t=1$ Myrs is shown for $q<4$ orbits with $a>18400$ au ({\it solid blue}) and with $10^4<a<18400$ au ({\it solid orange}). Orbits are weighted by their orbital frequency. The $s_{2\omega}$ distribution is also shown for LPCs making $q<4$ au perihelion passages 2--3 Myrs after HD 7977's encounter, weighted using Equation \ref{eq:whip}. Distributions are shown for LPCs with $a>10^4$ au ({\it dotted blue}) and with $10^3<a<10^4$ au ({\it dotted orange}). {\bf B:} For each of our HD 7977 passage simulations, the fraction of $q<4$ au orbits with $s_{2\omega}<0$ at $t=1$ Myrs is plotted against HD 7977's impact parameter. These fractions are calculated for orbits with $a>18400$ au ({\it blue}) and with $10^4<a<18400$ au ({\it orange}). Dashed lines mark the fractions for observed dynamically new LPCs ({\it blue}) and observed young returning LPCs ({\it orange}). {\bf C:} At $t=1$ Myr, the relative fractions of $q<4$ au orbits with $a>18400$ au ({\it blue}) and with $10^4<a<18400$ au ({\it orange}) are plotted against HD 7977's impact parameter.  }
\label{fig:assiso}
\end{figure}

The Flyby\_7943 simulation featured in Figure \ref{fig:orbperiod} provides our best match to observed LPC $s_{2\omega}$ distributions. One may expect similar results for other HD 7977 impact parameters, but this is often not what we find. The match to observations is degraded when we explore both closer and further impact parameters for HD 7977. The reason for this can be seen in Figure \ref{fig:assiso}. In Panel A, we examine an orbital snapshot of our Flyby\_7943 simulation at $t=1$ Myrs, and we split all bodies with perihelion under 4 au into two groups: 1) those with semimajor axes between 10$^4$ and 18400 au and 2) those with semimajor axis over 18400 au. Since $a=18400$ au corresponds to a 2.5-Myr orbital period, comets in the first group will make a perihelion passage before $t=2.5$ Myrs (which we take as equivalent to the current epoch) and can become young returning LPCs. Meanwhile, comets making their first perihelion passage at $t=2.5$ Myrs or later have to belong to the second group of bodies. Figure \ref{fig:assiso}A shows that at $t=1$ Myr, these two groups have distinctly different $s_{2\omega}$ distributions. Furthermore, these $s_{2\omega}$ distributions at $t=1$ Myr are similar to the distributions that we gather from simulated LPCs making actual perihelion passages in the $t=$ 2--3 Myr window. (One can see that the $t=1$ Myr distribution for $a>18400$ au bodies is actually slightly more isotropic than the one compiled from perihelion passages between 2--3 Myrs after HD 7977. This is because some of the orbits with $s_{2\omega}<0$ at $t=1$ Myr move back out of the inner solar system before they actually undergo a perihelion passage.)

In Figure \ref{fig:assiso}B, we analyze the $s_{2\omega}$ distributions of these two groups of bodies at $t=1$ Myr for all of our other passage simulations (Flyby\_3162 through Flyby\_50119). As a proxy to gauge the isotropy of a given $s_{2\omega}$ distribution, we consider the fraction of orbits with $s_{2\omega}<0$. We see in fact that these fractions most closely match the observed LPC $s_{2\omega}$ distributions for the Flyby\_7943 simulation. If we instead consider the Flyby\_6310 simulation, we still find a disparity between bodies with $a>18400$ au and bodies with 10$^4<a<18400$ au, but it is smaller than Flyby\_7943. In this case, HD 7977's passage is not only strong enough to inject a large number of LPCs with $a<18400$ au into the inner solar system, but it also isotropizes their $s_{2\omega}$ distribution more than the Flyby\_7943 simulation. As a result, the difference between the $s_{2\omega}$ distributions of new and young returning LPCs is smaller in the aftermath of the HD 7977 passage. This isotropization of young returning LPCs becomes stronger and stronger as HD 7977 is assigned still closer impact parameters, which decreases $s_{2\omega}$ differences between new and young returning LPCs. 

On the other hand, the Flyby\_10000 simulation considers a more distant HD 7977 impact parameter of $10^4$ au. In this case, there is still a difference in the $s_{2\omega}$ distributions of LPCs that come from closer and from more distant parts of the Oort cloud. However, two other effects occur. First, this weaker stellar encounter is not quite strong enough to isotropize $a>18400$ au orbits to the level seen in Flyby\_7943, and the $s_{2\omega}<0$ fraction falls below the observed level. The second effect can be seen in Figure \ref{fig:assiso}C. The weaker encounter simply fails to inject as many LPCs into the inner solar system on orbits with $a<18400$ au. Thus, a greater fraction of the young returning LPC population 2--3 Myrs after the HD 7977 passage is derived from the $a>18400$ au population, which also supplies most of the population of new LPCs after the passage, and the two $s_{2\omega}$ distributions become more similar (and less isotropic). For impact parameters beyond 10$^4$ au (simulations Flyby\_12539--Flyby\_50119), the fraction of $a<18400$ au orbits continues to fall relative to those with $a>18400$ au, forcing young returners to share nearly the same parent orbital population as new LPCs. 

\section{Discussion}\label{sec:dis}

\subsection{Semimajor Axis Discrepancy}

\begin{figure}
\centering
\includegraphics[scale=0.55]{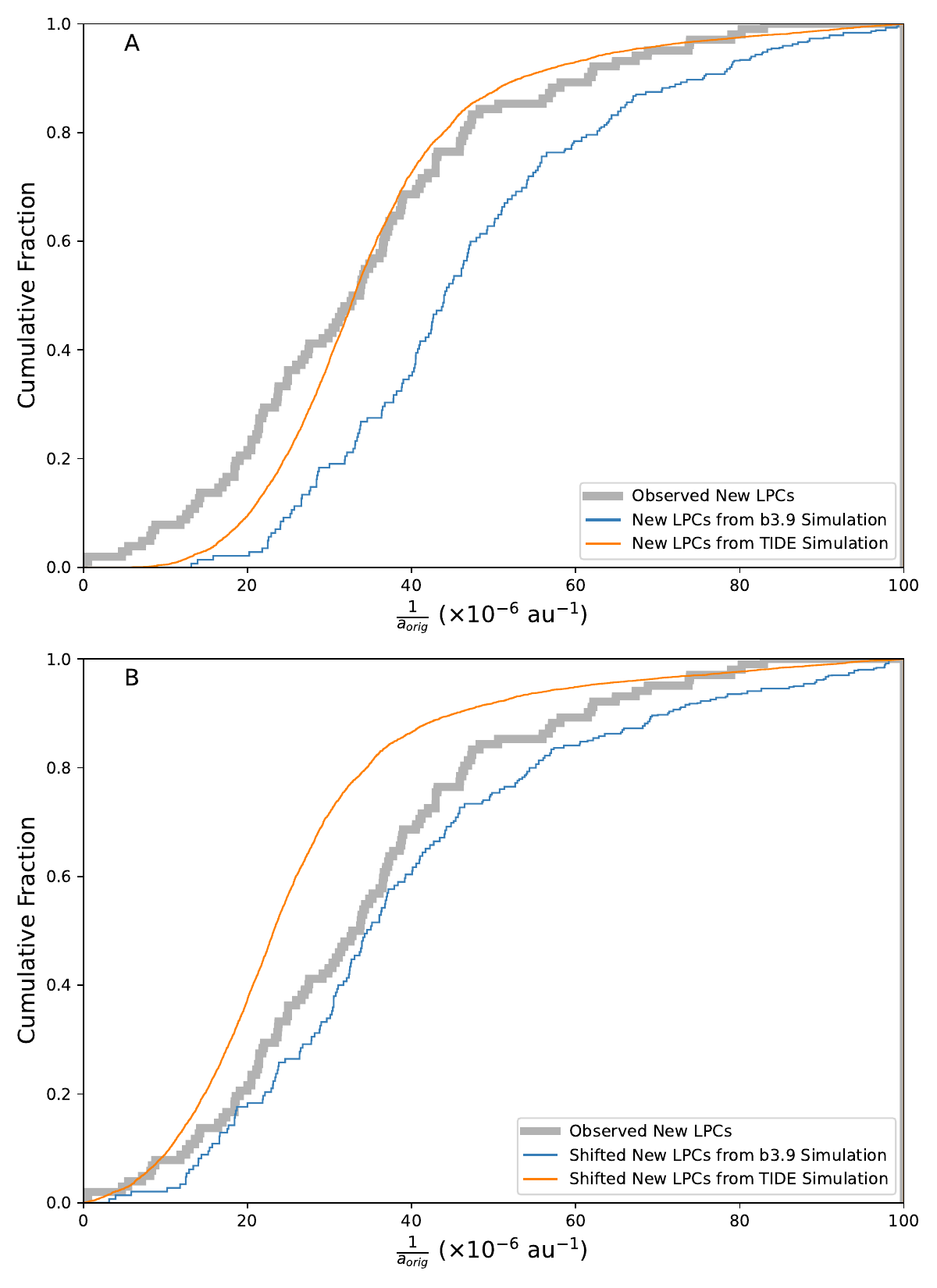}
\caption{{\bf A:} Distribution of inverse original semimajor axes of dynamically new ($a>10^4$ au) LPCs is shown for the TIDE ({\it orange}) and Flyby\_7943 ({\it blue}) simulations (weighted with Equation \ref{eq:whip}). The observed distribution using LPCs falling in the shaded region of Figure \ref{fig:obstrends} is also shown in gray. {\bf B:} Same as panel A, except the simulated distributions are shifted toward 0 by 10$^{-5}$ au$^{-1}$.}
\label{fig:smacompare}
\end{figure}

Until this point, the work presented here focuses almost exclusively on the distributions of $s_{2\omega}$ for LPCs. We now direct our attention to the semimajor axis distributions of LPCs. In Figure \ref{fig:smacompare}A, we plot the observed distribution of the inverse of original semimajor axes for all the dynamically new LPCs falling in the shaded region of Figure \ref{fig:obstrends}. Although the CODE catalog provides preferred orbits for some LPCs that account for non-gravitational forces, this is not done for all comets, so here we only consider orbital fits from gravitational models in the CODE catalog, which are provided for all of the catalog's comets\footnote{In instances where more than one gravitational modeling orbit is provided for the same comet in the CODE catalog, we average the different models.}. (We should note that a typical LPC's $s_{2\omega}$ variations are negligible when including or excluding non-gravitational forces.) In addition, we also plot the distributions from our TIDE simulation as well as our Flyby\_7943 simulation. Here we see that the original semimajor axes of the Flyby\_7943 simulation provide a poor match to the observed distribution. A K-S test comparing the observed distribution with the Flyby\_7943 simulation returns a $p$-value of just $\sim$10$^{-7}$. Meanwhile, the same comparison between  the TIDE simulation and observations yields $p=0.012$, indicating that the TIDE simulation produces a better match but that the null hypothesis can still be rejected with greater than 2$\sigma$ confidence. 

Although Figure \ref{fig:smacompare}A appears to suggest that our Flyby\_7943 simulation strongly conflicts with the observed semimajor axes of new LPCs, there may be less confidence in this conclusion than the plot seems to imply. Since LPC orbital energies are so small, the non-gravitational forces that cometary activity drives can have significant impacts on our measurements of LPC original semimajor axes, unlike values of $s_{2\omega}$. \citet{mars73} first noted that accounting for cometary non-gravitational acceleration in LPC observations seems to result in orbital solutions that imply a tighter original semimajor axis. Using the same non-gravitational modeling formalism, \citet{kro20} found that consideration of non-gravitational accelerations leads to a typical shift of $\sim$10$^{-5}$ au$^{-1}$ in the original LPC semimajor axis toward tighter binding energies compared to purely gravitational fitting. For instance, if purely gravitational fitting of an observed comet yields an original semimajor axis is $3\times10^4$ au, properly accounting for non-gravitational forces may imply the actual original semimajor axis is $\sim$$2.3\times10^4$ au.) In light of this result, in Figure \ref{fig:smacompare}B, we attempt to crudely bias our simulated distributions to account for the effects of non-gravitational influence that are absent in our simulations but are known to operate among the observed LPCs. To implement this crude bias, we simply shift our simulated orbits by 10$^{-5}$ au$^{-1}$ to larger semimajor axis. After doing this, we find that the Flyby\_7943 simulation matches the observations much better. A K-S test now returns a $p$-value of 0.16. Meanwhile, the same comparison using the TIDE simulation returns a $p$-value of $4\times10^{-8}$. Again, this simple shift we apply is likely a gross oversimplification of the effects of non-gravitational forces, and \citet{krodon23} find that a systematic shift to more tightly bound original orbits may be unique to the relatively simple parameterization of ice sublimation employed in \citet{mars73}. Nevertheless, it is clear that non-gravitational forces do significantly alter measurements of LPC original semimajor axes \citep{krodon23}. The simulated orbital energy shift we implement is not meant to confirm the validity of a particular simulation, but rather to demonstrate that the unaltered observed original semimajor axis distribution of dynamically new LPCs may not confidently rule out either LPC simulation.

\subsection{Oort Cloud Structure Uncertainty}

Another possible interpretation of the semimajor axis discrepancy in Figure \ref{fig:smacompare}A is that the Oort cloud semimajor axis distribution we employ in our simulations is simply different from the real Oort cloud's distribution. The Oort cloud's semimajor distribution is very poorly constrained by LPC orbits, and our understanding of it is almost completely model-dependent. Although the Oort spike of new LPCs resides near $a\sim3\times10^4$ au, this does not necessarily mean that the Oort cloud extends out to $3\times10^4$ au. The reason for this is that as the perihelia of Oort cloud bodies with smaller semimajor axes ($5\times10^3 \lesssim a \lesssim 2\times10^4$ au) are slowly driven Sunward, perturbations from Jupiter and Saturn can inflate their semimajor axes to larger values ($a\gtrsim3\times10^4$ au) by the time we observe them entering the inner solar system as ``new'' LPCs \citep{kaibquinn09}. On the other hand, LPCs observed in the inner solar system do not strictly require Oort cloud bodies on orbits with $a\lesssim3\times10^4$ au, either. The perihelia of Oort cloud bodies with larger semimajor axes ($a\gtrsim3\times10^4$ au) evolve much more rapidly each orbital revolution compared to those on tighter orbits, and they can slide past the giant planets into the inner solar system without having their semimajor axes heavily modified during prior perihelion passages \citep{heitre86}. 

\begin{figure}
\centering
\includegraphics[scale=0.35]{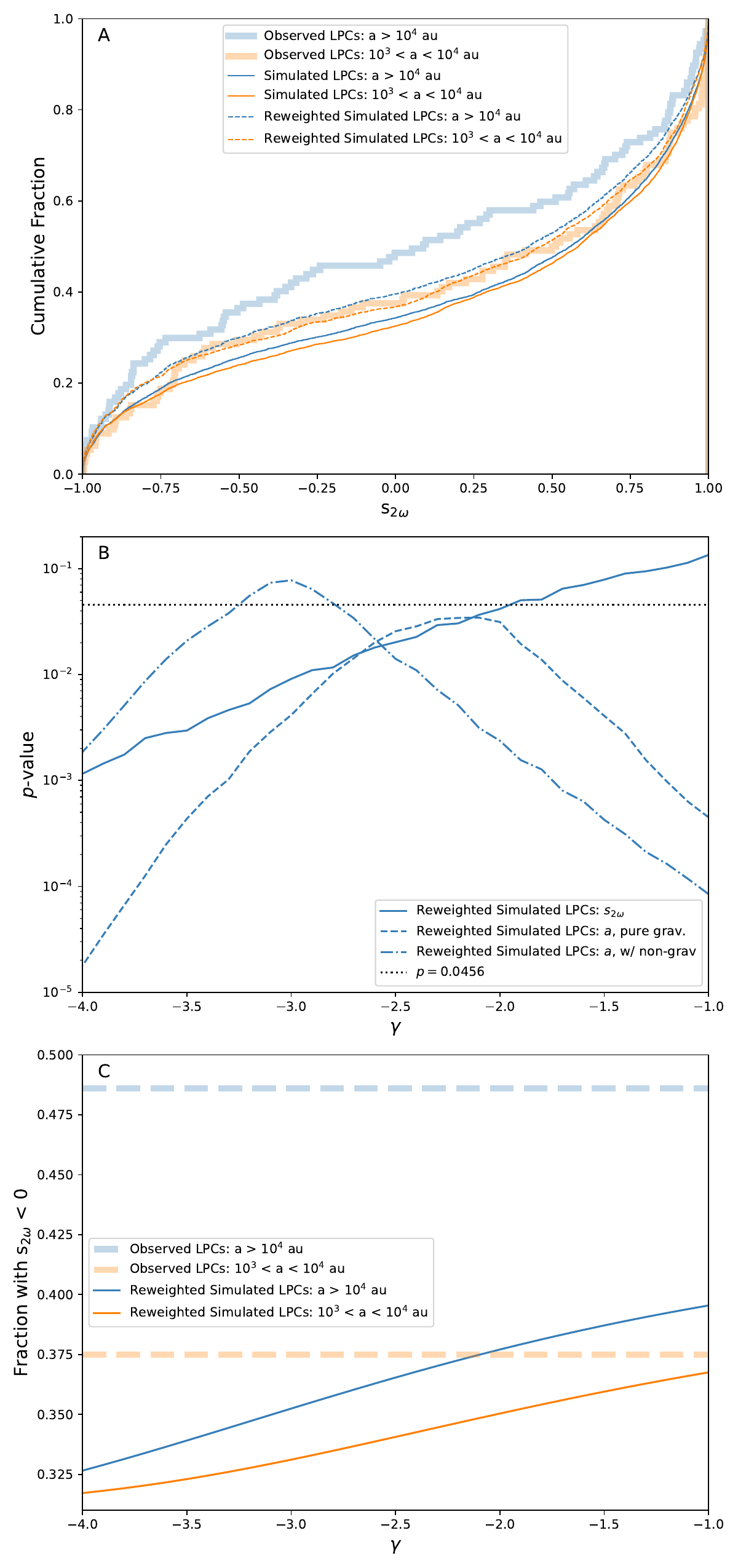}
\caption{{\bf A:} The cumulative distribution of $s_{2\omega}$ is shown for observed LPCs ({\it thick lines}) and simulated LPCs from the TIDE simulation ({\it thin lines}). LPCs are split into dynamically new comets ($a>10^4$ au; {\it blue}) and returning comets with $10^3<a<10^4$ au ({\it orange}). For the TIDE simulation we also show distributions if we reweight our results assuming an initial density profile proportional to $a^{-1}$ ({\it dashed lines}). All simulated LPCs are also weighted according their fading probability given in Equation \ref{eq:whip}.  {\bf B:} The $p$-value comparing $s_{2\omega}$ values ({\it solid blue}) of observed new comets with simulated new comets from reweighted TIDE simulations is plotted against the value of $\gamma$ ($n\propto a^\gamma$) assumed in our reweighting. We also plot the $p$-values comparing original semimajor axes of observed and simulated new comets. We perform comparisons to fitted observed comet orbits that consider non-gravitational forces ({\it dash-dotted blue}) and that exclude non-gravitational forces ({\it dashed blue}). {\bf C:} The fraction of LPCs with $s_{2\omega}<0$ is plotted against the $\gamma$ value used in simulation reweighting. Fractions are shown for new LPCs ({\it blue}) and young returning LPCs ({\it orange}). Horizontal dashed lines mark the observed fractions. }
\label{fig:altocdist}
\end{figure}

The main motivation for our simulations' initial power-law density profile of $n(a)\propto a^{-3.35}$ is that it is similar to the predictions of numerous simulations that model the formation of the Oort cloud from the planetary scattering of leftover planetesimals in the giant planet region \citep[e.g.][]{dqt87, kaibquinn08, vok19}. However, given the very weak constraints on the true profile, we now assess how much the discrepancy between the TIDE simulation's $s_{2\omega}$ distributions and those observed for LPCs depends on our chosen semimajor axis distribution. In Figure \ref{fig:altocdist}A, we illustrate how the TIDE simulation's $s_{2\omega}$ distribution for new LPCs changes when we reweight our simulated orbits to radically shift the effective density power-law index from -3.35 to -1. We see in this figure that the change to the simulated $s_{2\omega}$ distributions is modest but also toward better agreement with observations. For simulated LPCs with $a>10^4$ au and $q<4$ au, the fraction of comets with negative $s_{2\omega}$ values increases from 0.343 to 0.395. While this reweighted distribution still appears less isotropic than observed new LPCs, it is no longer statistically rejectable, as a K-S test returns a $p$-value of 0.13. 

In Figure \ref{fig:altocdist}B, we plot the $p$-value returned by K-S tests comparing simulated and observed new LPCs ($a>10^4$ au) as a function of $\gamma$, the power-law indices we assume in the initial density distributions of reweighted TIDE simulations. We see that for $\gamma$ values below $\sim$-2, the null hypothesis can be rejected with at least $2\sigma$ confidence. However, this is not so for $\gamma\gtrsim-1.9$. In Figure \ref{fig:altocdist}C, we plot the fraction of LPCs with $s_{2\omega}<0$ as a function of $\gamma$. This is done for both new LPCs ($a>10^4$ au) and young returning LPCs ($10^3<a<10^4$ au). We see that no matter what our assumed semimajor axis distribution, the observed difference the new and young returning $s_{2\omega}$ distributions is never close to replicated in the TIDE simulation. Instead, in the non-rejectable cases, the simulated $s_{2\omega}$ distributions fall close enough to the average of the two observed distributions that the observed differences can be chalked up to fluctuations from small number statistics. 

The reason the fraction of comets $s_{2\omega}<0$ goes up with increasing $\gamma$ is because this flattens the Oort cloud's density profile, and bodies with initial semimajor axes beyond $\sim$$3\times10^4$ au contribute more to the population of new LPCs. In our unweighted run ($\gamma=-3.35$), only 25\% of our simulated new LPCs began the TIDE simulation with $a>3\times10^4$ au. When we reweight particles to mimic $\gamma=-3.35$, this fraction increases from 25\% to 53\%. The $s_{2\omega}$ values of LPCs with such large semimajor axes fluctuate more rapidly under the Galactic tide, and this increases the chance that they can flip from positive (perihelia moving Sunward) to negative (perihelia moving away from the Sun) prior to making their first perihelion passage through the inner solar system. 

Thus, if the underlying semimajor axis distribution is much less centrally concentrated than our TIDE simulation assumes, it may be that there is no conflict between simulated LPC production and the observed $s_{2\omega}$ distributions for new and young returning LPCs. This flat density profile is difficult to reconcile with the canonical model of Oort cloud formation \citep[e.g.][]{dqt87}, but the alternative formation scenario of Oort cloud capture during stellar birth cluster dissolution could generate a less centrally concentrated cloud \citep{lev10}. However, this capture scenario would imply no correlation between Oort cloud structure and the Invariable Plane, but there are hints from comet orbit statistics that the Oort cloud began as a flattened structure \citep{hig20, fouch23}. 

Of course, flattening out the Oort cloud's density profile will obviously have an effect on the distribution of original semimajor axes that simulated new LPCs possess when they make their first perihelion passage through the inner solar system. Consequently, we can also compare the distributions of simulated, reweighted new LPC original semimajor axes to their observed counterparts. Using a K-S test for this comparison, we plot the $p$-values in Figure \ref{fig:altocdist}B. These comparisons are done against the pure gravitational orbital fits in the CODE catalog as well as the ``preferred'' CODE orbits, which may include non-gravitational fits. In the case of comparisons with observed orbits that can include non-gravitational forces, simulated original semimajor axes are not rejectable at the 2$\sigma$ level for $\gamma$ between -3.25 and -2.8, which is far from the $\gamma$ range that generates non-rejectable $s_{2\omega}$ distributions. If we instead consider pure gravitational fits to observed comets that are no simulated distributions of original semimajor axes that are non-rejectable. The highest $p$-value attained is $\sim$0.03 near $\gamma=-2.2$. For larger values of $\gamma$, the match to observed original semimajor axes becomes worse and worse, and there is no value of $\gamma$ that simultaneously replicates the observed distributions of original semimajor axes and $s_{2\omega}$.

\subsection{Comet Fading Uncertainty}

Throughout most of this work, we employ a relatively simple and widely used form of comet survival probability to account for the fact that comets seem to fade from observability over a smaller number of perihelion passages than dynamical models predict \citep{whip62, vok19}. While finding an optimal fading law through a full parameter sweep of various fading law families is beyond the scope of this work, we explore here how our simulations' ability to match observations can be influenced by our fading law choice. Our chosen \citet{whip62} fading assumes that comet survival probability declines via a smooth power-law as a comet's number of perihelion passages inside some critical distance ($r=4$ au in our work) increases. However, this is far from the only possibility. For instance, \citet{weis80} suggests bimodal fading behavior, where most comets fade rapidly within the first few passages through the inner solar system but a smaller fraction survive much longer. To explore this alternative fading behavior, we formulate a rather extreme version of it to gauge its potential effects; we assume that 90\% of LPCs fade after just a single passage through the inner 4 au of the solar system, and the remaining 10\% never fade and are only removed via dynamical means (typically ejection).

\begin{figure}
\centering
\includegraphics[scale=0.55]{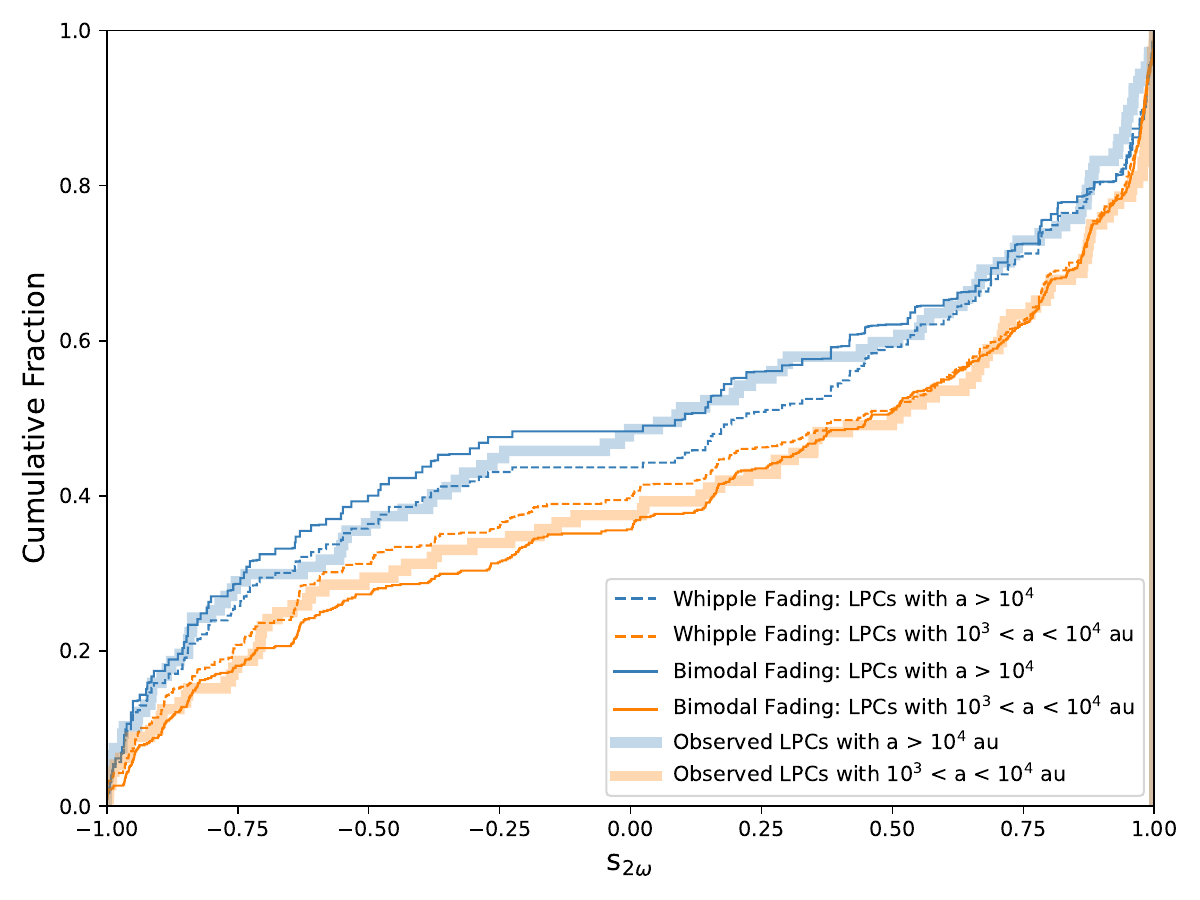}
\caption{Distributions of $s_{2\omega}$ are shown for LPCs making perihelion passages inside 4 au with $a>10^4$ au ({\it blue}) and $10^3<a<10^4$ au ({\it orange}). Observed distributions ({\it thick solid lines}) are compared against simulated LPCs from Flyby\_6310 that make perihelion passages 2--3 Myrs after the HD 7977 flyby. Two different comet fading laws are assumed: a Whipple-like fading ({\it dashed}) and a bimodal fading ({\it thin solid}).} 
\label{fig:altfade}
\end{figure}

In Figure \ref{fig:altfade}, we show how both assumed fading laws affect the outcome of our Flyby\_6310 simulation, analyzing simulated LPCs that pass through the inner 4 au of the solar system 2--3 Myrs after HD 7977's $\sim$6300-au passage. Here we see that the \citet{whip62} fading law generates a more isotropic $s_{2\omega}$ distribution for new LPCs than for young returning ones. However, the differences between the two distributions are not as distinct as in the Flyby\_7943 simulation featured in Figure \ref{fig:orbperiod}. For Flyby\_6310, the simulated new LPCs are not isotropic enough, and the simulated young returning LPCs are too isotropic compared to the observed distributions. Figure \ref{fig:altfade} shows that this can change if we consider other LPC fading scenarios, such as bimodal fading. In the bimodal fading case, the simulated new LPCs and young returning LPCs now exhibit $s_{2\omega}$ distributions that are actually slightly more disparate than the observed LPCs. One reason for this is that the bimodal fading strongly suppresses any contribution that earlier epochs of LPCs provide to the LPCs with $a>10^4$ au in the 2--3 Myr window. LPCs injected in these earlier epochs tend to come from more interior regions of the Oort cloud with $s_{2\omega}$ distributions biased toward 1. Their suppression leads to a more isotropic population of new LPCs at $t=$ 2--3 Myrs. In addition, the bimodal fading law means that all LPCs with $10^3<a<10^4$ au in the 2--3 Myr window receive equal weighting, since comet survival probability flatlines after the first passage through the inner solar system, and these comets are making their second or later perihelion passage. This equal weighting biases our simulated young returning LPCs at 2--3 Myrs even more strongly toward earlier epochs where the $s_{2\omega}$ distribution is less isotropic. In other words, our alternative fading scheme enhances the differences in the initial times of entry into the inner solar system between new and returning LPCs. 

Figure \ref{fig:altfade} shows that our assumed comet fading law can enhance or suppress the differences in the $s_{2\omega}$ distributions for new and young returning LPCs. Nevertheless, there is a limit to this imposed by the actual post-flyby orbital distributions. For instance, Figure \ref{fig:assiso} shows that for impact parameters closer than 5000 au (Flyby\_3162 thru Flyby\_5012 simulations), there is nearly no variation in the level of $s_{2\omega}$ isotropy with semimajor axes beyond 10$^4$ au. Thus, it is not clear than any fading law can generate large differences between young returning LPCs and new LPCs in these cases. In addition, for impact parameters beyond 10$^4$ au (Flyby\_12539 thru Flyby\_50119 simulations), the level of $s_{2\omega}$ isotropy in semimajor axes over 18400 au (with orbital periods over 2.5 Myrs) is well below the levels observed among new LPCs and alternative fading laws cannot push the simulated LPCs nearer to this observed isotropy.

\subsection{Implications for Oort Cloud Properties}

If the disparity in the observed $s_{2\omega}$ distributions is in fact caused by a close flyby of HD 7977, this means that our inner solar system has in fact experienced a comet shower in the last $\sim$2.5 Myrs \citep{hills81, cao26}. Although the bulk of the LPCs injected into the inner solar system during a comet shower come to perihelion within $\sim$1 Myr, our passage simulations show that the LPC production rate remains elevated for 3+ Myrs after HD 7977's passage, and modest LPC production enhancements may persist for tens of Myrs after a strong stellar passages \citep{fouch11}. One reason for this is that, outside of periods of strong stellar perturbation, the Oort cloud's loss cone ($q\lesssim15$ au) is only truly filled for semimajor axes beyond $\sim$50000 au \citep{rick08}. For smaller semimajor axes between 30000--50000 au, tide-dominated simulations show that the number of comets with $q<15$ au is significantly less than what would be expected from a fully thermalized population \citep{rick08}. However, a strong perturbation from HD 7977 will temporarily greatly enhance the number of comets with perihelia in the loss cone for a wide range of semimajor axes. As our Figure \ref{fig:orbperiod} shows, those LPCs with small semimajor axes ($a\sim10^4$ au) arrive in the first Myr after the stellar passage, and the loss cone for this semimajor axis range empties back out on this timescale (the LPCs' orbital periods). However, for semimajor axes of 20000--50000 au, the new HD 7977-provided additions to the loss cone will not be fully removed for 3--10 Myrs after the stellar encounter (again, the range of orbital periods). Furthermore, there can be a ``second wave'' of LPCs in the inner solar system from bodies with $a\lesssim10^4$ au whose perihelia are initially injected into the Jupiter-Saturn region ($q=$ 5--15 au). A subset of these bodies can have their semimajor axes inflated to larger values from the planetary energy kicks incurred during their first post-HD 7977 perihelion passage, after which the Galactic tide can then inject them further into the inner solar system on their next revolution about the Sun \citep{kaibquinn09}. One can see several examples of these second wave LPCs on the right hand side of Figure \ref{fig:orbperiod}A below the dashed line in the panel. Thus, the effects of a strong stellar passage on the new LPC flux can persist for several Myrs after the passage. 

\begin{figure}
\centering
\includegraphics[scale=0.55]{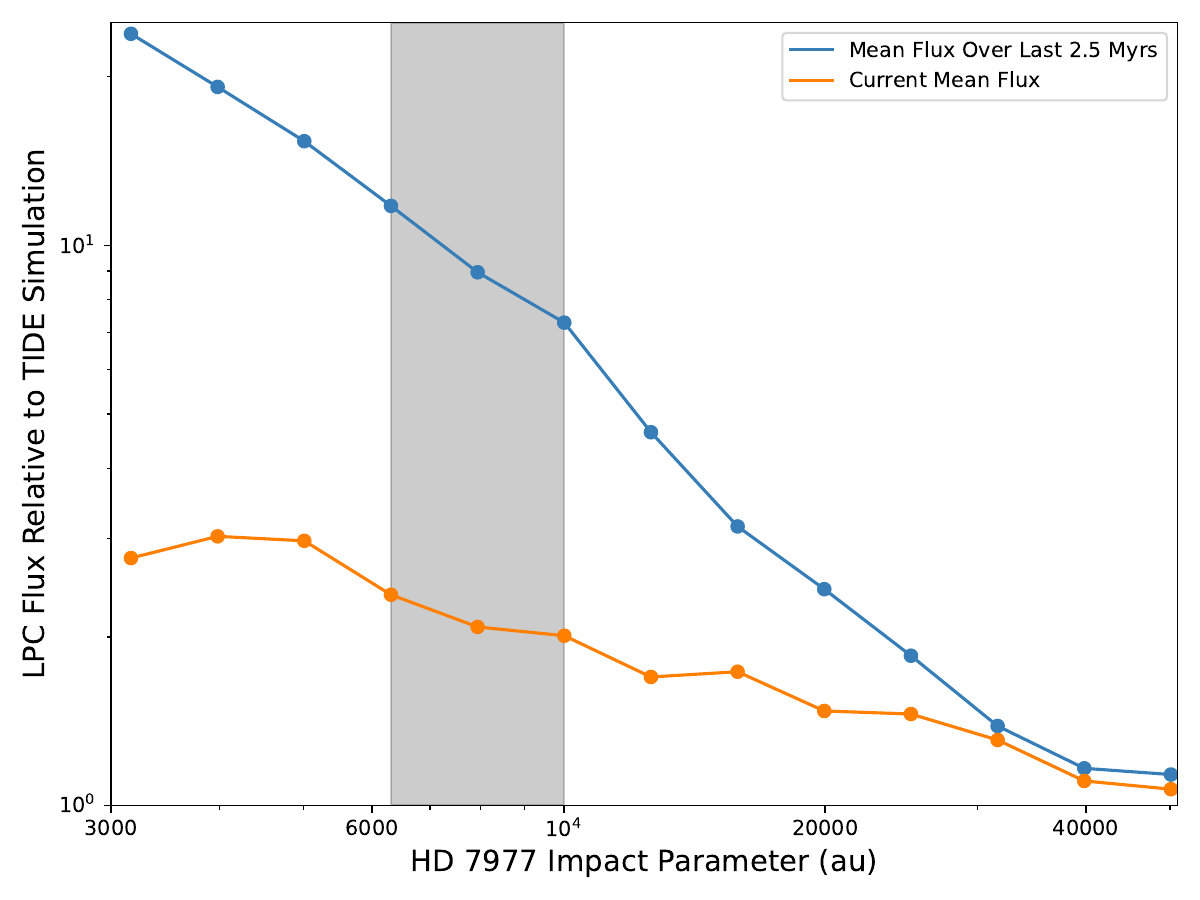}
\caption{Plot of the mean new LPC flux into the inner solar system against the impact parameter of HD 7977 used in our different passage simulations. Mean LPC flux is averaged over the first 2.5 Myrs after the passage ({\it blue}) and 2--3 Myrs after the passage ({\it orange}). }
\label{fig:showpow}
\end{figure}

We have shown that only HD 7977 impact parameters between $\sim$6000--10000 au appear capable of replicating the heightened $s_{2\omega}$ isotropy of observed new LPCs relative to young returning LPCs. This range of impact parameters places constraints on how powerful a comet shower related to HD 7977 could have been. In Figure \ref{fig:showpow}, we plot the flux of new LPCs into the inner solar system over the last 2.5 Myrs against the impact parameter used for HD 7977 in each of our passage simulations. This flux is plotted as a factor of the expected flux derived from our TIDE simulation. As can be seen, for stellar passages between 6000--10000 au, the inner solar system would have been exposed to $\sim$7--12 times as many LPCs during the past 2.5 Myrs as would be expected during periods of Oort cloud dynamics dominated by the Galactic tide. Figure \ref{fig:showpow} also shows that even our weaker encounters produce a slight enhancement of LPCs, with the mean LPC flux over the last 2.5 Myrs being elevated by 10--20\% for the two rightmost data points. Measured by solar impulse or impulse gradient, stellar encounters of these strengths should be expected every $\sim$5--10 Myrs on average, which suggests that LPC flux can vary by, perhaps, a couple of tens of percent on $\sim$Myr timescales.

It is important to note that it takes longer than 2--3 Myrs after HD 7977's passage for the LPC flux to return to the levels expected from our TIDE simulation. In Figure \ref{fig:showpow}, we also plot the mean flux of new LPCs (relative to the TIDE flux) during the 2--3 Myrs after each HD 7977 passage. Here we find that LPC flux is still elevated by a factor of $\sim$2--2.5 relative to the TIDE simulation for impact parameters between 6000 and 10000 au. This means that even today, the observed flux of LPCs in the inner solar system may be roughly double the value it has been for most of the solar system's history. This may alleviate the finding of \citet{vok19}, who calculated that their simulated Oort cloud (which did not include a recent strong stellar perturbation) failed by a factor of $\sim$2--3 to replicate the observed flux of comets if the LPC size distribution of \citet{sosfern11} was adopted. Thus, the elevated LPC flux implied by our simulations has an important consequence for the Oort Cloud's population size. The number of bodies in the Oort cloud is typically inferred through comparisons of the observed flux of dynamically new LPCs in the inner solar system to the flux generated in simulations of Oort cloud LPC generation, wherein the Galactic tide has provided the largest recent perturbations on the Oort cloud \citep[e.g.,][]{wietre99, kaibquinn09}. If HD 7977 is responsible for the $s_{2\omega}$ isotropy observed in new LPCs, then prior estimates of the Oort cloud's population are likely a factor of 2--2.5 too large. 

\section{Summary \& Conclusions}\label{sec:con}

To assess the dominance of the Galactic tide in the recent production of LPCs, we study the distribution of $s_{2\omega}$ among LPCs. We find that observed new LPCs in the inner solar system ($q<4$ au; $a>10^4$ au) have a more isotropic distribution of $s_{2\omega}$ than new LPCs from an Oort cloud simulation where the Galactic tide dominates perturbations on the Oort cloud. However, the $s_{2\omega}$ distribution of observed young returning LPCs ($q<4$ au; $10^3<a<10^4$ au) does display anisotropy consistent with Galactic tidal perturbation.

In addition to our Oort cloud simulation dominated by the Galactic tide, we also explore how the modern LPC population can be altered by HD 7977, a Sun-like star thought to have passed between 4000--24000 au from the Sun $\sim$2.47--2.76 Myrs ago \citep{bail22, dyb24}. To do this, we perform 13 other simulations modeling the production of LPCs in the $\sim$3 Myrs after HD 7977 passage with varying impact parameters from 3200--51000 au. We find that only HD 7977 passages within 6000--10000 au can simultaneously replicate a nearly isotropic $s_{2\omega}$ distribution for new LPCs and a significantly less isotropic one for young returning LPCs. The reason is that such passages are strong enough to inject a large number of LPCs with $a\simeq10^4$ au into the inner solar system, and strong enough to completely isotropize the $s_{2\omega}$ distribution for LPCs with $a\gtrsim2\times10^4$ au, but also weak enough for newly injected LPCs with $a\simeq10^4$ au to retain some of their tide-related $s_{2\omega}$ anisotropy. In these encounter scenarios, the $s_{2\omega}$ isotropy increases with LPC semimajor axis, and due to their shorter orbital periods, the less isotropic LPCs arrive in the inner solar system more quickly than the more isotropic LPCs after HD 7977's passage. These less isotropic, early-arriving LPCs are the main progenitors of the population of young returning LPCs present 2--3 Myrs after the HD 7977 passage, and the young returners therefore retain an anisotropic $s_{2\omega}$ distribution. Meanwhile, owing to their longer orbital periods, many of the HD 7977-injected LPCs with $a\gtrsim2\times10^4$ au do not arrive in the inner solar system until 2--3 Myrs after the HD 7977 passage. Their nearly isotropic $s_{2\omega}$ values give rise to a nearly isotropic population of new LPCs that is observed concurrently with a less isotropic population of young returning LPCs. 

Our HD 7977 passage simulations that best replicate the observed $s_{2\omega}$ distributions of LPCs notably fail to replicate the observed orbital energies of dynamically new LPCs. However, this may not be a fatal flaw for these models, as it is well-established that non-gravitational forces alter the inferred original semimajor axes of LPCs \citep{mars73}. If these non-gravitational forces typically result in pure gravitational fits systematically overestimating LPC orbital energies by $\sim$10$^{-5}$ au$^{-1}$ \citep{kro20}, then our simulation results compare well with observed LPCs. With longer observing timelines and larger heliocentric distances of detection, LPC observations with the Vera Rubin Observatory may be able to better constrain the degree and direction of any systematic orbital shift from non-gravitational forces \citep{ivez19}. 

Another possibility is that our failure to replicate the observed orbital energies of dynamically new LPCs indicates that the underlying semimajor axis distribution we assume for our simulated Oort cloud is incorrect. Reweighting our tide-dominated simulation particles to achieve a flatter Oort cloud density profile, we find that the simulation's $s_{2\omega}$ distribution for new LPCs becomes more isotropic as the contribution of large semimajor axis orbits to LPC production is enhanced. However, these flatter density profiles for the Oort cloud are in conflict with the tradition model of Oort cloud formation that invokes the capture of material scattered by the giant planets \citep{dqt87}. Moreover, these flatter density profiles still cannot simultaneous reproduce the observed distributions of $a$ and $s_{2\omega}$ for new LPCs. 

Our work here shows that modern LPC orbits can be used to infer the power of the solar system's past few Myrs of stellar passages, and these orbits suggest a past encounter distance for HD 7977 of 6000--10000 au from the Sun. This implies we are still living through the late stages of a comet shower that has elevated the flux of LPCs through the inner solar system by an order of magnitude since HD 7977's passage. Even today, we would expect the flux to be $\sim$twice as high as epochs where the Galactic tide solely dominates LPC production. A consequence of this is that estimates of the number of Oort cloud bodies based on the modern flux of LPCs should likely be revised downward by a factor of two or more. Our LPC orbital analysis predicts that the next Gaia Data Release \citep{gaiaDR3} will favor an HD 7977 encounter distance of $\sim$6000--10000 au. If our prediction is confirmed, this would likely mean HD 7977's passage was the most powerful stellar encounter the solar system has experienced in the past 100--300 Myrs. 

\section{Acknowledgements}

NAK's contributions to this work were supported from NASA Solar System Workings grant 80NSSC24K1874. SNR is grateful to the CNRS/INSU's PNP program and to the French Space Agency (CNES) for support. We thank Ma\l gorzata Kr\'{o}likowska-So\l tan for valuable conversations regarding non-gravitational accelerations of LPCs.

\bibliography{LPCFunniness}

\end{document}